\documentclass[useAMS,usenatbib]{mn2e}
\usepackage{wrapfig}
\usepackage{multicol}
\usepackage{epsfig}

\title[Protostellar discs formed from rigidly rotating cores]
{Protostellar discs formed from rigidly rotating cores}
\author[S.~Walch, A.~Burkert, A.~Whitworth, T.~Naab, M.~Gritschneder]
{S. Walch$^{1,2}$\thanks{E-mail: Stefanie.Walch@astro.cf.ac.uk}, 
A.~Burkert$^{1}$, A.~Whitworth$^{2}$, T.~Naab$^{1}$, M.~Gritschneder$^{1}$\\
$^{1}$University Observatory, University of Munich, Scheinerstr. 1, Munich, 81679, Germany \\
$^{2}$School of Physics \& Astronomy, Cardiff University, 5 The Parade, Cardiff CF24 3AA, Wales, UK}
\begin{document}

\date{Accepted . Received 2009 January 13; in original form }

\pagerange{\pageref{firstpage}--\pageref{lastpage}} \pubyear{2009}

\maketitle

\label{firstpage}

\begin{abstract}
We use three-dimensional SPH simulations to investigate the collapse of low-mass prestellar cores and the formation and early evolution of protostellar discs. The initial conditions are slightly supercritical Bonnor-Ebert spheres in rigid rotation. The core mass and initial radius are held fixed at $M_{_{\rm O}}=6.1\,{\rm M}_\odot$ and $R_{_{\rm O}}=17,000\,{\rm AU}$, and the only parameter that we vary is the initial angular speed $\Omega_{_{\rm O}}$. Protostellar discs forming from cores with $\Omega_{_{\rm O}}<1.35\times 10^{-13}\,{\rm s}^{-1}$ have radii between 100 and $300\,{\rm AU}$ and are quite centrally concentrated; due to heating by gas infall onto the disc and accretion onto the central object, they are also quite warm, $\bar{T}>100\,{\rm K}$, and therefore stable against gravitational fragmentation. In contrast, more rapidly rotating cores form discs which are less concentrated and cooler, and have radii between 400 and $1000\,{\rm AU}$; as a consequence they are prone to gravitational fragmentation and the formation of multiple systems. We derive a criterion that predicts whether a rigidly rotating core having given $M_{_{\rm O}}$, $R_{_{\rm O}}$ and $\Omega_{_{\rm O}}$ will produce a protostellar disc which fragments whilst material is still infalling from the core envelope. We then apply this criterion to core samples for which $M_{_{\rm O}}$, $R_{_{\rm O}}$ and $\Omega_{_{\rm O}}$ have been estimated observationally. We conclude that the observed cores are stable against fragmentation at this stage, due to their low angular speeds and the heat delivered at the accretion shock where the infalling material hits the disc.
\end{abstract}



\section{Introduction}

Stars form from the gravitational collapse of prestellar cores. We expect that, to a first approximation, angular momentum is conserved minutely during the early stages of collapse, and this leads to the formation of a protostellar disc and/or to fragmentation into a multiple system. Thus, understanding the formation and evolution of protostellar discs is a critical stage linking the initial conditions within prestellar cores to the final outcome of the star formation process.

In this paper, we simulate the collapse of isolated, rigidly rotating cores, using the 3D OpenMP-parallelised SPH code VINE \citep{Wetzstein2008, Nelson2008}. All cores have the same mass and the same initial density profile, given by a Bonnor-Ebert sphere (BES) \citep{Ebert1955, Bonnor1956}; they are distinguished only by their angular momentum. We study the influence of core angular momentum on the properties of the protostellar discs and protostars that condense out of such discs. In a companion paper (Walch et al., in prep.), we explore and contrast the properties of protostellar discs and protostars condensing out of turbulent cores having the same net angular momenta as the rigidly rotating cores considered here.

Previous numerical work on the collapse and fragmentation of cores has demonstrated that the critical ingredients are the net amount of angular momentum and its distribution, the initial density profile, and the treatment of thermodynamics \citep[for a summary see ][]{Goodwin2007PPV}. For example, \citet{Miyama1984} and \citet{Burkert1993} examine the evolution of cores with a uniform initial density and an isothermal equation of state. They find that the outcome depends on the parameters $\alpha\equiv U_{_{\rm THERM}}/|U_{_{\rm GRAV}}|$ and $\beta\equiv U_{_{\rm ROT}}/|U_{_{\rm GRAV}}|$, where $U_{_{\rm THERM}}$, $U_{_{\rm GRAV}}$ and $U_{_{\rm ROT}}$ are the thermal, gravitational and rotational energies of the initial core. Specifically, \citet{Miyama1984} only find fragmentation for $\alpha\beta<0.12$. 


However, observations of prestellar cores suggest that they are centrally condensed, and well represented by a Bonnor-Ebert profile \citep{Alves2001, Andre2004}, which is the equilibrium configuration for an isothermal cloud. Centrally condensed clouds appear to be more stable against fragmentation than cores with a uniform density \citep{Lynden-Bell1964, Boss1993, Burkert1996}. For a review see \citet{Bodenheimer2000}.

To date, the only parameter studies of the collapse of Bonnor-Ebert spheres have been performed using grid codes. \citet{Matsumoto2003} investigate the collapse of $3\,{\rm M}_\odot$ critical Bonnor-Ebert spheres with the density enhanced by 10\%. They invoke a barotropic equation of state, and investigate different total angular momenta, different rotation laws, and bar mode perturbations of different strengths. Their simulations use a nested grid with mirror symmetry about the equatorial plane, and fine resolution in the centre (the location of the primary protostar), but quite coarse resolution in the outer parts (the disc region), particularly in the vertical direction. They conclude that fragmentation requires $\Omega_{_{\rm C}}t_{_{\rm FF}}>0.05$, where $\Omega_{_{\rm C}}$ and $t_{_{\rm FF}}$ are the initial angular speed and freefall time at the centre of the core; this is equivalent to a lower limit on $\beta$. The study of disc fragmentation by \citet{Nelson2006} suggests that these results may be compromised by poor vertical resolution. 

\citet{Banerjee2004} investigate the collapse of a $2.1\,{\rm M}_\odot$ core and a $168\,{\rm M}_\odot$ core, using the FLASH AMR code. The cores are initially Bonnor-Ebert spheres with the density enhanced everywhere by 10\%, and then a 10\% $\,m\!=\!2$ bar perturbation superimposed; they are initially in rigid rotation. In order to capture the thermodynamics more realistically they use the tabulated molecular-line cooling prescription of \citet{Neufeld1995}. They perform only one simulation of a low mass core, which has $\Omega_{_{\rm C}}t_{_{\rm FF}}=0.2$ and produces a bar; they suggest that this bar might subsequently fragment if the simulation were followed further.

\citet{Boss2000} have demonstrated that core fragmentation is highly sensitive to the treatment of thermodynamics. However, although \citet{Krumholz2006} have performed 3D AMR simulations of $100\,{\rm M}_\odot$ and $200\,{\rm M}_\odot$ collapsing cores with radiative transfer, there has as yet been no comprehensive parameter study of low-mass collapsing cores using a realistic treatment of the gas thermodynamics. This paper is an attempt to perform such a study.

In Sections 2 and 3 we describe the initial conditions, the numerical code and the constitutive physics. In Sections 4 and 5 we present in detail the evolution of cores with, respectively, low and high $\Omega_{_{\rm O}}$, and in Section 6 we describe how the evolution changes as $\Omega_{_{\rm O}}$ is increased between these limits. In the Appendix we develop an analytic model, and derive a condition for fragmentation based on the initial parameters of a core. In Section 7 this criterion is applied to a sample of observed cores, and it is shown that they are all stable against disc fragmentation. In Section 8 we summarise our main conclusions.

\section{Initial Conditions}\label{initial}

\subsection{The initial core density profile}\label{BES}

The observed column-density profiles of isolated prestellar cores are often well matched by approximately critical Bonnor-Ebert spheres, i.e. equilibrium isothermal gas clouds which are close to the critical state for gravitational condensation \citep[see e.g. ][]{Alves2001, Hennebelle2003}. We have therefore modeled our initial cores as marginally supercritical Bonnor-Ebert spheres, truncated at dimensionless radius $\xi_{_{\rm B}}=6.9$ (the critical value is $\xi_{_{\rm B}}=6.45$) and then with their density increased everywhere by 10\%, to ensure immediate contraction. The gas is assumed to be molecular hydrogen at $28\,{\rm K}$, and therefore the isothermal sound speed is $a_{_{\rm O}}=0.34\,{\rm km}\,{\rm s}^{-1}$. The central density is $\rho_{_{\rm C}}=10^{-18}\,{\rm g}\,{\rm cm}^{-3}$, and so the mass and radius are
\begin{eqnarray}
M_{_{\rm O}}\!&\!=\!&\!1.1\,\frac{a_{_{\rm O}}^3}{G\left(4\pi G\rho_{_{\rm C}}\right)^{1/2}}\,\left\{\xi_{_{\rm B}}^2\psi'_{_{\rm B}}\right\}\,\;=\;\,6.1\,{\rm M}_\odot\,,\\
R_{_{\rm O}}&=&\frac{a_{_{\rm O}}}{\left(4\pi G\rho_{_{\rm C}}\right)^{1/2}}\,\xi_{_{\rm B}}\,\;=\;\,17,000\,{\rm AU}\,.
\end{eqnarray}
Here $\psi(\xi)$ is the Isothermal Function and $\psi'(\xi)$ its first derivative \citep[see][]{Chandrasekhar1949}; $\psi_{_{\rm B}}\equiv\psi\left(\xi_{_{\rm B}}\right)$ and $\psi_{_{\rm B}}'\equiv\psi'\left(\xi_{_{\rm B}}\right)$. The core must be contained by an external pressure
\begin{eqnarray}\label{EQN:PEXT}
P_{_{\rm EXT}}&=&1.1\,\rho_{_{\rm C}}\,a_{_{\rm O}}^2\,{\rm e}^{-\psi_{_{\rm B}}}\;=\;k_{_{\rm B}}\,5.5\times 10^5\,{\rm cm}^{-3}\,{\rm K}\,,
\end{eqnarray}
delivered by a hot rarefied medium. The freefall time at the core centre (hence the timescale on which the central primary protostar starts to condense out) is $t_{_{\rm FF}}=67\,{\rm kyr}$. The thermal energy of the core is $U_{_{\rm THERM}}=3M_{_{\rm O}}a_{_{\rm O}}^2/2$, and the self-gravitational potential energy is
\begin{eqnarray}\nonumber
U_{_{\rm GRAV}}\!\!&\!\!=\!\!&\!\!-\frac{3}{\xi_{_{\rm B}}\psi'_{_{\rm B}}}\left\{1-\frac{\xi_{_{\rm B}}{\rm e}^{-\psi_{_{\rm B}}}}{3\psi'_{_{\rm B}}}\right\}\frac{GM_{_{\rm O}}^2}{R_{_{\rm O}}}\,=\,0.74\,\frac{GM_{_{\rm O}}^2}{R_{_{\rm O}}}.
\end{eqnarray}
Hence the ratio of thermal to gravitational energy is
\begin{eqnarray}
\alpha&\equiv&\frac{U_{_{\rm THERM}}}{|U_{_{\rm GRAV}}|}\;=\;2.0\;\,\frac{a_{_{\rm O}}^2\,R_{_{\rm O}}}{G\,M_{_{\rm O}}}\;=\;0.74\,;
\end{eqnarray}
we note that $\alpha > 1/2$, because of the external pressure term in the Virial Theorem.

\subsection{The initial velocity field}

The only core property which we vary is the initial angular speed, $\Omega_{_{\rm O}}$. For a core in rigid rotation about the unit vector $\hat{\bf k}$, the initial velocity field is then
\begin{eqnarray}
{\bf v}({\bf r})&=&\Omega_{_{\rm O}}\,\hat{\bf k}\wedge{\bf r}\,.
\end{eqnarray}
The moment of inertia of the core is
\begin{eqnarray}
I_{_{\rm O}}&=&0.28\,M_{_{\rm O}}\,R_{_{\rm O}}^2\,.
\end{eqnarray}
Hence the total specific angular momentum is given by
\begin{eqnarray}
j_{_{\rm O}}&=&1.8\times 10^{21}\,{\rm cm}^2\,{\rm s}^{-1}\,\left(\frac{\Omega_{_{\rm O}}}{10^{-13}\,{\rm s}^{-1}}\right)\,,
\end{eqnarray}
and the initial ratio of rotational to gravitational energy by
\begin{eqnarray}
\beta&=&0.19\;\,\frac{R_{_{\rm O}}^3\,\Omega_{_{\rm O}}^2}{G\,M_{_{\rm O}}}\;\,\simeq\;\,0.036\,\left(\frac{\Omega_{_{\rm O}}}{10^{-13}\,{\rm s}^{-1}}\right)^2\,.
\end{eqnarray}

Our choice of representative values of $\Omega_{_{\rm O}}$ is informed by two constraints. (i) The total specific angular momenta in low-mass cores are typically of order $10^{21}\,{\rm cm}^2\,{\rm s}^{-1}$ \citep{Goodman1993}. (ii) The non-thermal velocity dispersions in low-mass cores are usually subsonic, and at most transonic \citep[e.g.][]{Barranco1998,Andre2007}. Therefore the rotational velocity at the edge of the core should not be much greater than $a_{_{\rm O}}$. The values of $\Omega_{_{\rm O}}$ that we consider are listed in Table \ref{TAB:SIMUS}; the equator of the fastest spinning core rotates at ${\rm Mach}\,1.5\,$.

\begin{table}
\begin{center}
\begin{tabular}{cccccccc}\hline
$\!\!{\rm Run}\!\!$ & $\Omega_{_{\rm O}}$ & $t_{_{\rm O}}$ & $t_{_{28}}$ & $M_\star$ & $M_{_{\rm D}}$ & $R_{_{\rm D}}$ & F \\
 & $\!\!\overline{10^{-13}{\rm s}^{-1}}\!\!$ & $\!\!\overline{\rm kyr}\!\!$ & $\!\!\overline{\rm kyr}\!\!$ & $\!\!\overline{{\rm M}_\odot}\!\!$ & $\!\!\overline{{\rm M}_\odot}\!\!$ & $\overline{\rm AU}$ & \\\hline
1 & 0.6 & 94.5 & 114.8 & 0.53 & 1.17 & 250 & N\\
2 & 0.8 & 94.5 & 115.4 & 0.31 & 1.39 & 344 & N \\
3 & 1.0 & 96.5 & 118.4 & 0.25 & 1.45 & 433 & N \\
4 & 1.2 & 96.5 & 119.2 & 0.23 & 1.47 & 567 & N \\
5 & 1.5 & 98.5 & 123.3 & 0.17 & 1.53 & 739 & Y \\
6 & 1.6 & $102.4\;\,$ & 126.7 & 0.24 & 1.46 & $1091\;$ & Y \\
7 & 1.8 & $104.4\;\,$ & 130.4 & 0.16 & 1.54 & 952 & Y \\\hline
\end{tabular}
\end{center}
\caption{Column 1 gives the ID of the simulation, and Column 2 gives the initial rotational frequency, $\Omega_{_{\rm O}}$. Column 3 gives the time at which the first protostellar object forms, $t_{_{\rm O}}$, and Column 4 gives the time at which 28\% of the mass is either in the protostellar disc or in protostars, $t_{_{28}}$. Column 5 gives the mass of the primary protostar at this juncture, $M_\star$, and Columns 6 and 7 give the mass, $M_{_{\rm D}}$, and radius, $R_{_{\rm D}}$, of the protostellar disc at the same time. Column 8 indicates whether the disc fragments.}\label{TAB:SIMUS}
\end{table}

\section[Numerical Method]{Numerical Method and constitutive physics}\label{method}

\subsection{The VINE code}

The simulations have been performed with the Tree-SPH code VINE. A full description of VINE can be found in \citet{Wetzstein2008} and \citet{Nelson2008}. VINE is parallelised with OpenMP directives. It invokes a leapfrog integrator, and individual particle time steps with a CFL tolerance parameter of 0.1. Gravitational accelerations are estimated using a tree, with opening angle $\theta=0.005$ \citep{Springel2001}. The gravitational softening length equals the hydrodynamical smoothing length \citep{BateBurkert97}, which is adapted so that each particle has ${\cal N}_{_{\rm NEIB}}=50\pm20$ neighbours. Hydrodynamical forces are treated with periodic boundary conditions, but gravitational forces are not. Artificial viscosity is treated using the standard prescription of \citet{Gingold1983} with $\alpha_{_{\rm AV}}=1$ and $\beta_{_{\rm AV}}=2$, plus the Balsara switch \citep{Balsara1995}. We have experimented with time-varying viscosity \citep{Morris1997}, but find that it makes no significant difference.

\subsection{Equation of state and molecular line cooling}\label{cool}

For the purpose of calculating the gross thermodynamics, we assume that the gas is pure molecular hydrogen, with ratio of specific heats $\gamma=7/5$ and molecular weight $\mu=2$; thus for simplicity we are neglecting the fact that the rotational degrees of freedom of H$_{_2}$ are not fully  excited at low temperatures, and we are neglecting the contribution of helium to $\mu$. The equation of state is 
\begin{eqnarray}
P&=&(\gamma-1)\,u\,,
\end{eqnarray}
and the isothermal sound speed is
\begin{eqnarray}
a&=&\left(\frac{k_{_{\rm B}}\,T}{\mu\,m_{\rm p}}\right)^{1/2}\,.
\end{eqnarray}

In solving the energy equation we adopt the following procedure. At low densities, $\rho<\rho_{_{\rm CRIT}}=10^{-13}\,{\rm g}\,{\rm cm}^{-3}$, we use the cooling rates computed by \citet{Neufeld1995}, with the constraint that the temperature is not allowed to fall below $T=9\,{\rm K}$. At high densities, $\rho>\rho_{_{\rm CRIT}}$, we switch off the radiative cooling, and the gas then evolves adiabatically. \citet{Neufeld1995} have computed the equilibrium abundances of key molecules and atoms (CO, H$_{_2}$O, H$_{_2}$, HCl, O$_{_2}$, C and O), and the resulting net cooling rate due to line emission, $\Lambda$, as a function of density and temperature, using the local velocity gradient method to treat optical-depth effects. Neufeld has kindly supplied these rates in the form of a look-up table for densities in the range $10^3\le\left(n(H_2)/{\rm cm}^{-3}\right)\le 10^{10}$ and temperatures in the range $10\le\left(T/{\rm K}\right)\le 2500$. Most previous work in this field has used a barotropic equation of state and therefore does not treat thermal inertia effects \citep[e.g.][]{Goodwin2004, Bate1998, Matsumoto2003}, although \citet{Banerjee2004} use the same procedure as us. By invoking line cooling we hope to capture more realistically the thermal behaviour of low-density material as it falls onto the protostellar disc and then spirals into the central protostar. Ideally radiative cooling should also take account of continuum emission from dust, in the regime where the gas and dust are thermally coupled, for example by using the algorithm developed by \citet{Stamatellos2007}. Unfortunately the computational machinery to treat both the line cooling regime and the dust cooling regime does not yet exist. We are currently developing an algorithm to combine these regimes.

\subsection{Resolution}

The core is modeled with 430,000 SPH particles, each having mass $m_{_{\rm SPH}}=1.4\times 10^{-5}\,{\rm M}_\odot$. Following \citet{BateBurkert97}, the minimum resolvable mass is therefore $2{\cal N}_{_{\rm NEIB}}m_{_{\rm SPH}}=1.4\times 10^{-3}\,{\rm M}_\odot$. For comparison the minimum possible Jeans mass (corresponding to $\rho=\rho_{_{\rm CRIT}}$ and $T=9\,{\rm K}$) is $1.5\times 10^{-3}\,{\rm M}_\odot$. Therefore the Jeans mass is always resolved. In addition there are 24,000 ambient particles exerting the external pressure that contains the core (see Eqn. \ref{EQN:PEXT}). These particles also have mass $m_{_{\rm SPH}}=1.4\times 10^{-5}\,{\rm M}_\odot$, but they have much higher temperature, and their density is adjusted so that they fill the space between the core and the periodic boundaries. We do not use sink particles. Instead we limit the CPU time by imposing a minimum smoothing length, $h_{_{\rm MIN}}=2\,{\rm AU}$. Consequently we are able to resolve the details of the protostellar disc, both radially and vertically, but not the central protostar. 

\subsection{Definitions}

Any SPH particle whose density rises above $\rho_\star\!=\!10^{-11}\,{\rm g}\,{\rm cm}^{-3}$ is presumed to be part of a protostar. The primary protostar is the one that forms first, at the centre of the core, and $t_{_{\rm O}}$ is its time of formation. Values of $t_{_{\rm O}}$ are given in Table \ref{TAB:SIMUS}. $\;t_{_{\rm O}}$ increases monotonically with $\Omega_{_{\rm O}}$, because more rapidly rotating cores take longer to reach the critical density, $\rho_\star$. We define the disc to be all material with density in the range $10^{-16}\,{\rm g}\,{\rm cm}^{-3}\leq\rho\leq 10^{-11}\,{\rm g}\,{\rm cm}^{-3}$, since in general this material has $|v_\phi| \gg |v_r|$ (where $v_\phi$ and $v_r$ are the azimuthal and radial components of velocity).

The evolution of a core can usefully be divided into two distinct phases. During the {\it Isothermal Collapse Phase} radiative cooling is very efficient and the gas temperature is roughly constant. Throughout most of the isothermal phase centrifugal acceleration is a secondary effect, and the core collapses on an approximately freefall time scale. The {\it Protostellar Phase} starts as soon as a disc forms. The pattern of evolution during the Protostellar Phase depends critically on $\Omega_{_{\rm O}}$. In cores with high $\Omega_{_{\rm O}}$, (a) formation of the primary protostar is delayed until a substantial disc has built up, and (b) the outer disc is prone to fragment producing secondary protstars. Since the Protostellar Phase is shorter than the Isothermal Collapse Phase, we give times in the form $t=t_{_{\rm O}}+\Delta t$, so that $\Delta t$ is time measured from the moment when the primary protostar forms.

Simulations are compared at time $t_{_{28}}$, which is defined to be when $28\%$ of the core mass (i.e. $\sim 1.7\,{\rm M}_\odot$) has density $\rho > 10^{-16}\,{\rm g}\,{\rm cm}^{-3}$, i.e. it is either in the disc or in a protostar. $\;28\%$ is the typical efficiency of star formation observed in cores \citep[e.g.][]{Motte1998, Alves2007}. In order to follow the simulations further, we would need to relax the imposition of adiabaticity at densities $\rho>10^{-13}\,{\rm g}\,{\rm cm}^{-3}$, so that the energy dissipation resulting from the redistribution of angular momentum can be radiated away. Without this radiation the long-term viscous evolution of the disc is inhibited.

\section{A low angular momentum core}\label{resultsI}

In this section we describe the evolution of a low angular momentum core, having $\Omega_{_{\rm O}}=0.6\times 10^{-13}\,{\rm s}^{-1}$, and hence $j_{_{\rm O}}=1.1\times 10^{21}\,{\rm cm}^2\,{\rm s}^{-1}$ and $\beta=0.014$. This is Run 1.

{\sc The Isothermal Collapse Phase.} Fig. 1 shows the density profile at various times during the Isothermal Collapse Phase of Run 1. The collapse is roughly spherically symmetric (at least until the late stages) and approximately self-similar: the central region with a flat density profile shrinks, both in radial extent and in mass. The infalling envelope with $\rho\propto |{\bf r}|^{-2}$ extends ever further inwards. Density profiles are computed by arranging the particles in order of increasing $|{\bf r}|$, then averaging the density of contiguous sets of 100 particles; evidently this works only as long as departures from spherical symmetry are small.

\begin{figure}
\label{FIG:RHOIP}
\psfig{figure=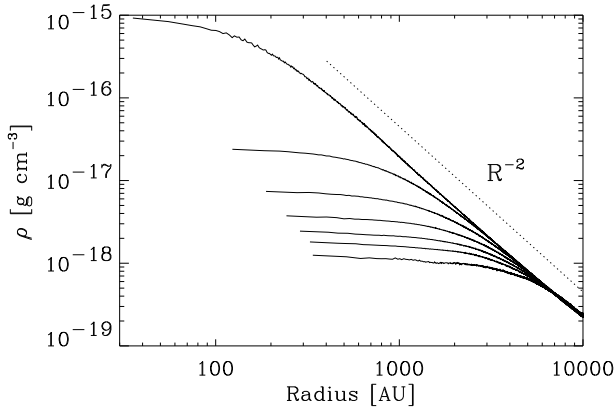, height=5.5cm}
\caption{Radial density profiles during the Isothermal Phase of Run 1. Reading from the bottom, profiles are shown at $t=0,\,41,\,51,\,61,\,71,\,81\;\&\;91\,{\rm kyr}$. The axes are logarithmic, and the dotted line gives a slope of $-\,2$.}
\end{figure}

{\sc The Protostellar Phase.} Fig. 2 shows velocity vectors superimposed on false-colour images of the density field on $400\,{\rm AU}\times 400\,{\rm AU}$ slices through the centre of the core during Run 1. These images correspond to times from just before the formation of the primary protostar at $t=t_{_{\rm O}}$ to $t=t_{_{28}}$. They show the formation of the primary protostar and its attendant protostellar disc. The density field is computed on a $64\times 64$ grid by first evaluating the column-density, $\Sigma$, through a slice having thickness $\Delta =10\,{\rm AU}$, and then setting $\rho=\Sigma/\Delta$. The velocity field is computed in an analogous way, but on a coarser $20\times 20$ grid.

\begin{figure}
  \epsfig{figure=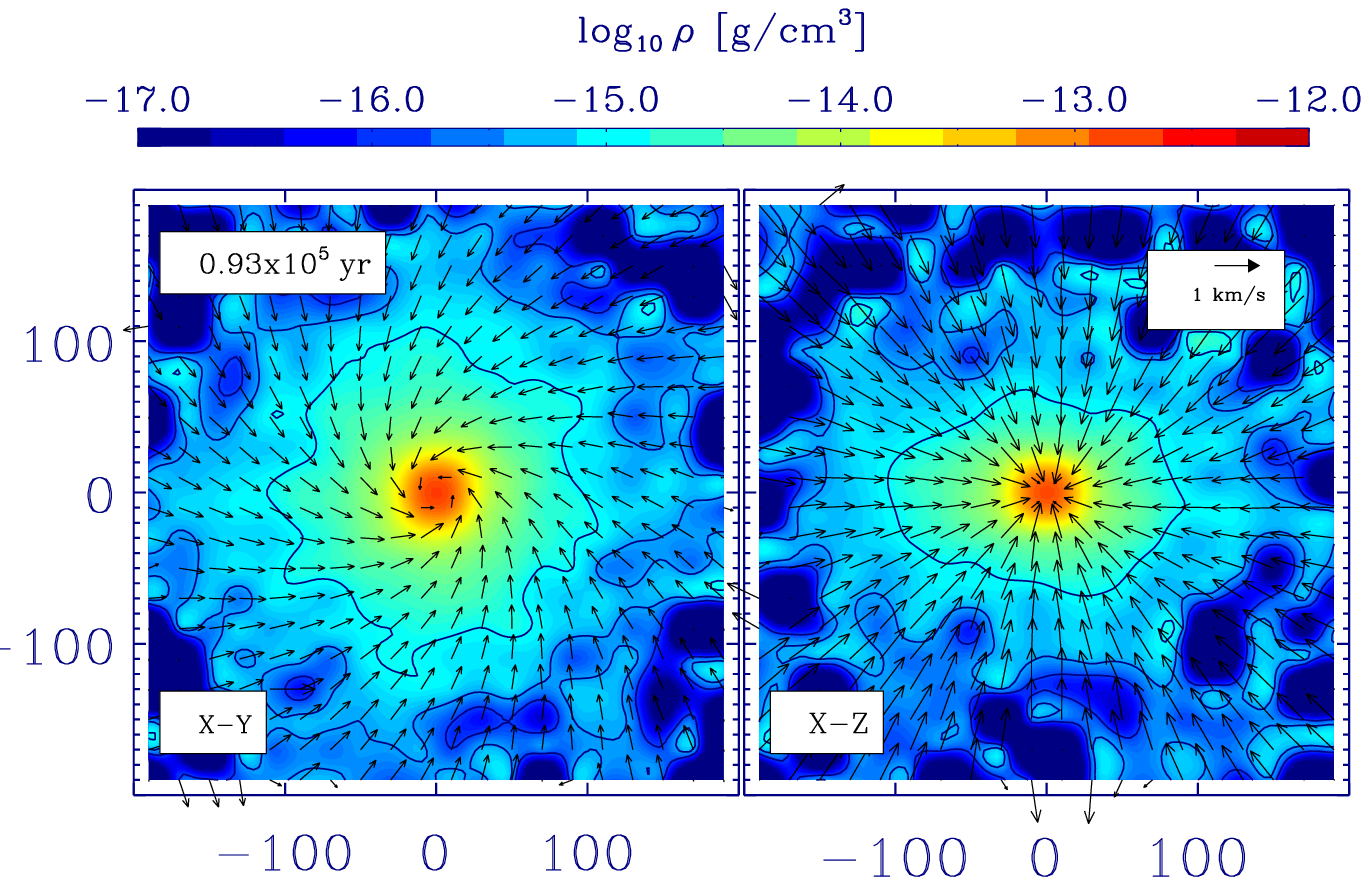, height=5.cm}
  \epsfig{figure=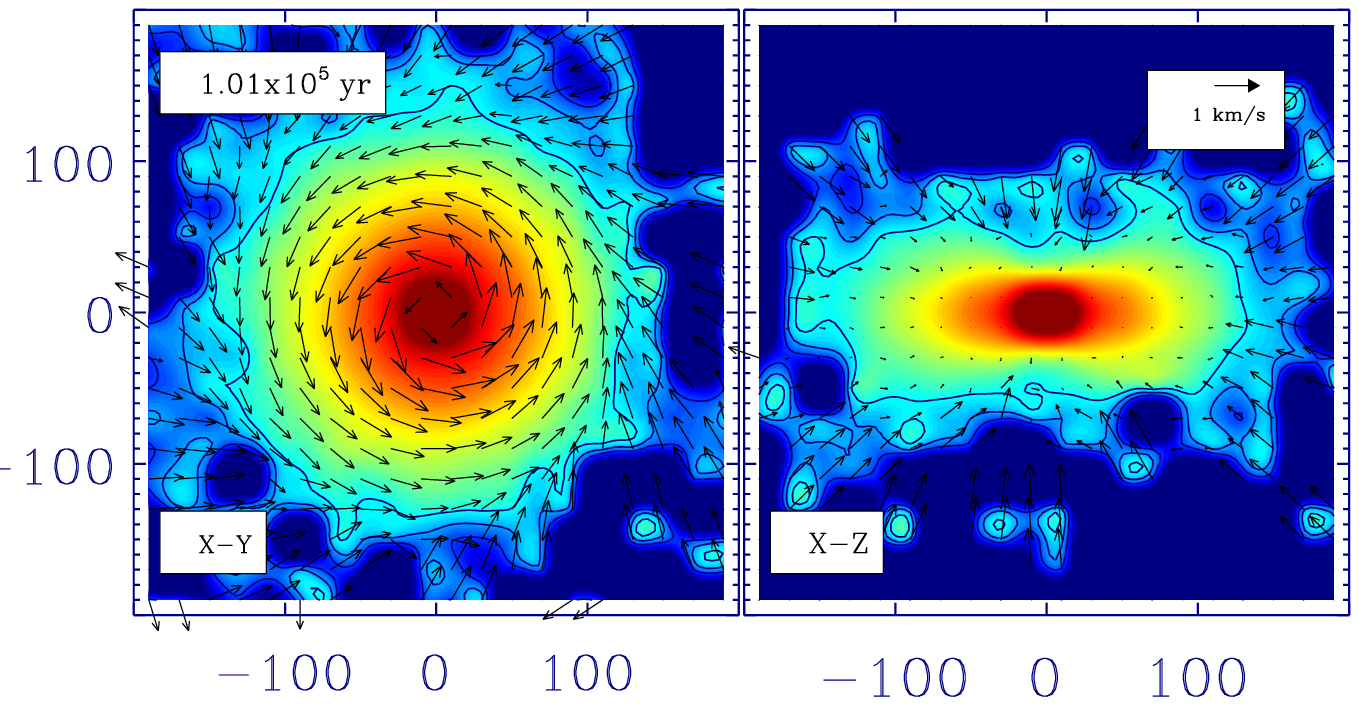, height=4.25cm}
  \epsfig{figure=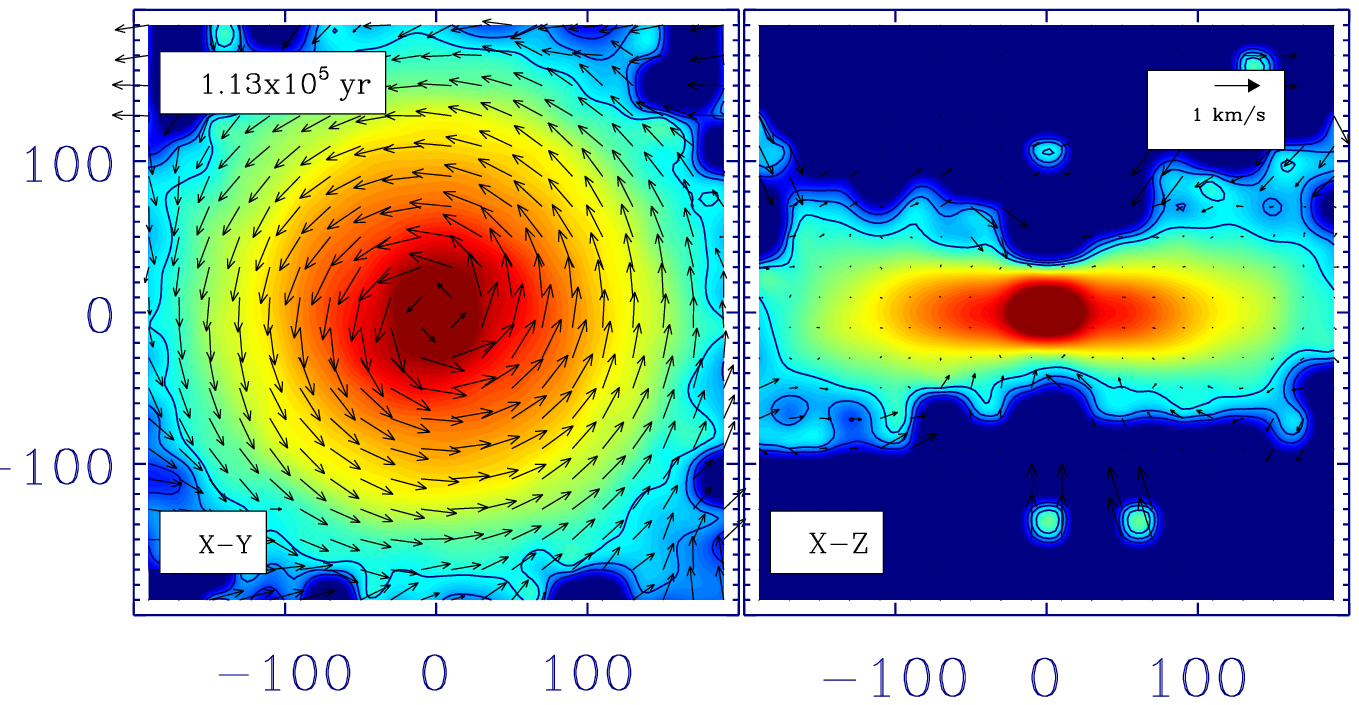, height=4.25cm}
\label{FIG:RUN1DVF}
\caption{The density on $400\,{\rm AU}\times 400\,{\rm AU}$ slices through the centre of the core during Run 1. The left-hand (respectively right-hand) column shows slices at $z\!=\!0\;\,\left({\rm respectively\;}y\!=\!0\right)$. The times shown are (a) $t_{_{\rm O}}-1.9\,{\rm kyr}$, (b) $t_{_{\rm O}}+6.2\,{\rm kyr}$, and (c) $t_{_{\rm O}}+18.2\,{\rm kyr}$. The density scale ranges from $10^{-18}\;{\rm to}\;10^{-13}\,{\rm g}\,{\rm cm}^{-3}$, and contours are plotted at $10^{-16},\,3\times 10^{-16}\;{\rm and}\;10^{-15}\,{\rm g}\,{\rm cm}^{-3}$. Velocity vectors are superimposed, and the insets show a $1\,{\rm km}\,{\rm s}^{-1}$ vector.}
\end{figure}

{\sc Density.} Fig. 3(a) shows radial density profiles on the equatorial plane, $\rho(r,z\!=\!0)$, from Run 1. Most of the mass interior to $10\,{\rm AU}$ is in the primary protostar. Outside this the disc extends to beyond $100\,{\rm AU}$, with $\rho\propto r^{-2.5}$. After $\,t=t_{_{\rm O}}+8.1\,{\rm kyr}$, the edge of the disc is very sharp.

{\sc Temperature.} Fig. 3(b) shows radial temperature profiles on the equatorial plane, $T(r,z\!=\!0)$, from Run 1. Throughout most of the disc, the run of temperature on the midplane approximates to $T\propto r^{-0.6}$. This is intermediate between the scaling expected for a thin, non--self-gravitating, steady-state $\alpha$-disc [which is heated by viscous dissipation, has a negative vertical temperature gradient, $dT/d|z|<0$, and $T_{_{\rm EFF}}\propto r^{-3/4}$ \citep{ShakuraSunyaev1973}] and the scaling inferred from observations of protoplanetary discs around classical T Tauri stars [which are mainly heated by stellar irradiation, have a positive vertical temperature gradient, $dT/d|z|>0$, and typically $T_{_{\rm EFF}}\propto r^{-1/2}$ \citep{Dullemond2001}]. Our simulations pertain to an earlier stage when the disc is self-gravitating and should be described as protostellar rather than protoplanetary. At this stage the infalling gas is heated impulsively when it encounters the accretion shock at the boundary of the disc; evidence for shock heating can be seen in the temperature spikes at the outer edge of the disc, on Fig. 3(b). Since the rate of infall onto the disc exceeds the rate of accretion through the disc and onto the primary protostar, the heating at the accretion shock dominates the energy budget of the disc. Subsequently the gas is heated further by compression and viscous dissipation as it spirals inwards and on to the primary protostar, and therefore the vertical temperature gradient is negative, $dT/d|z|<0$. Consequently we anticipate that, as a disc adjusts from being protostellar to being protoplanetary, the $10\,\mu{\rm m}$ silicate feature will switch from absorption to emission, as the sign of the vertical temperature gradient changes from negative to positive.
 
\begin{figure*}
 \begin{multicols}{2}
\psfig{figure=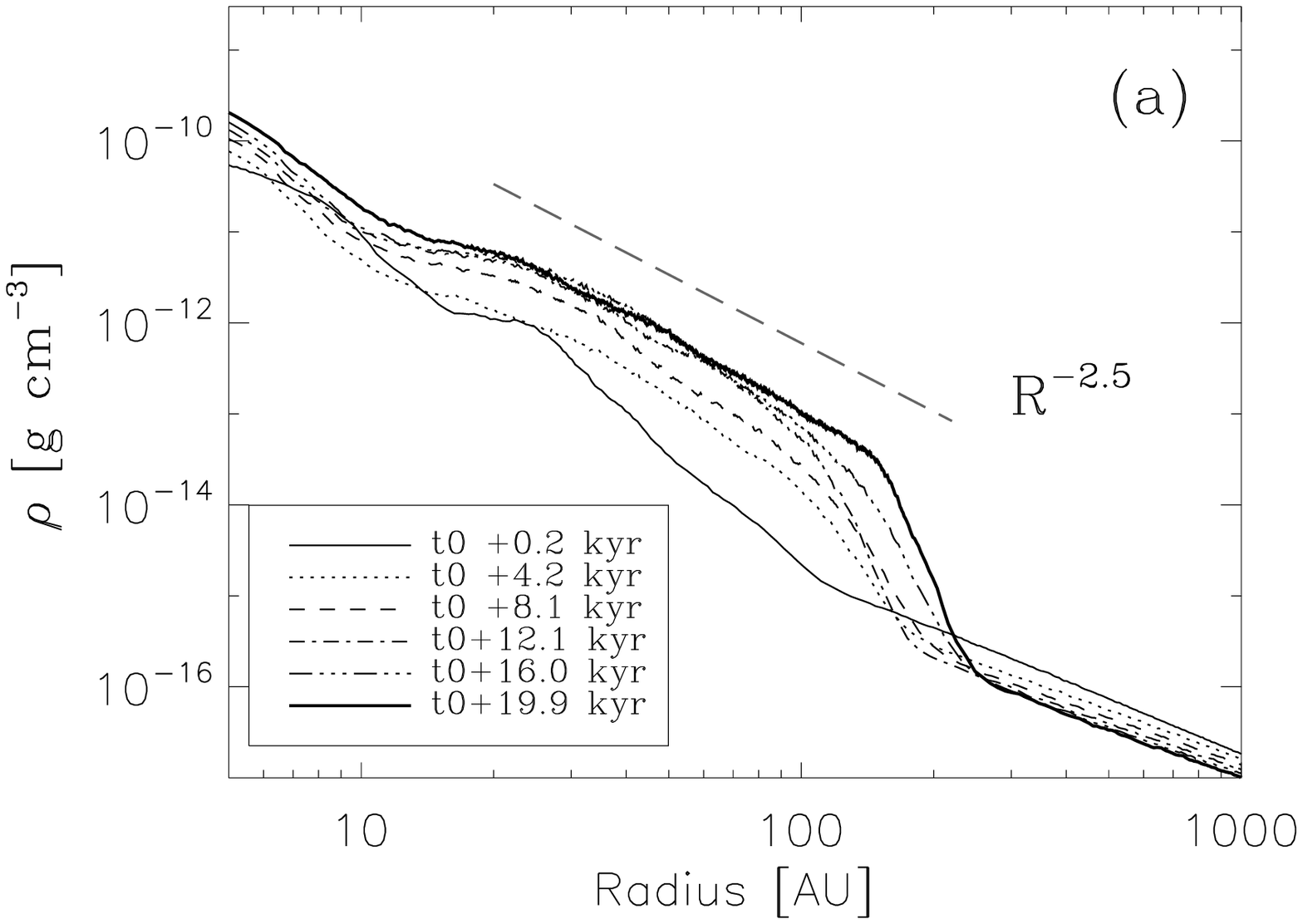, height=6cm}
\psfig{figure=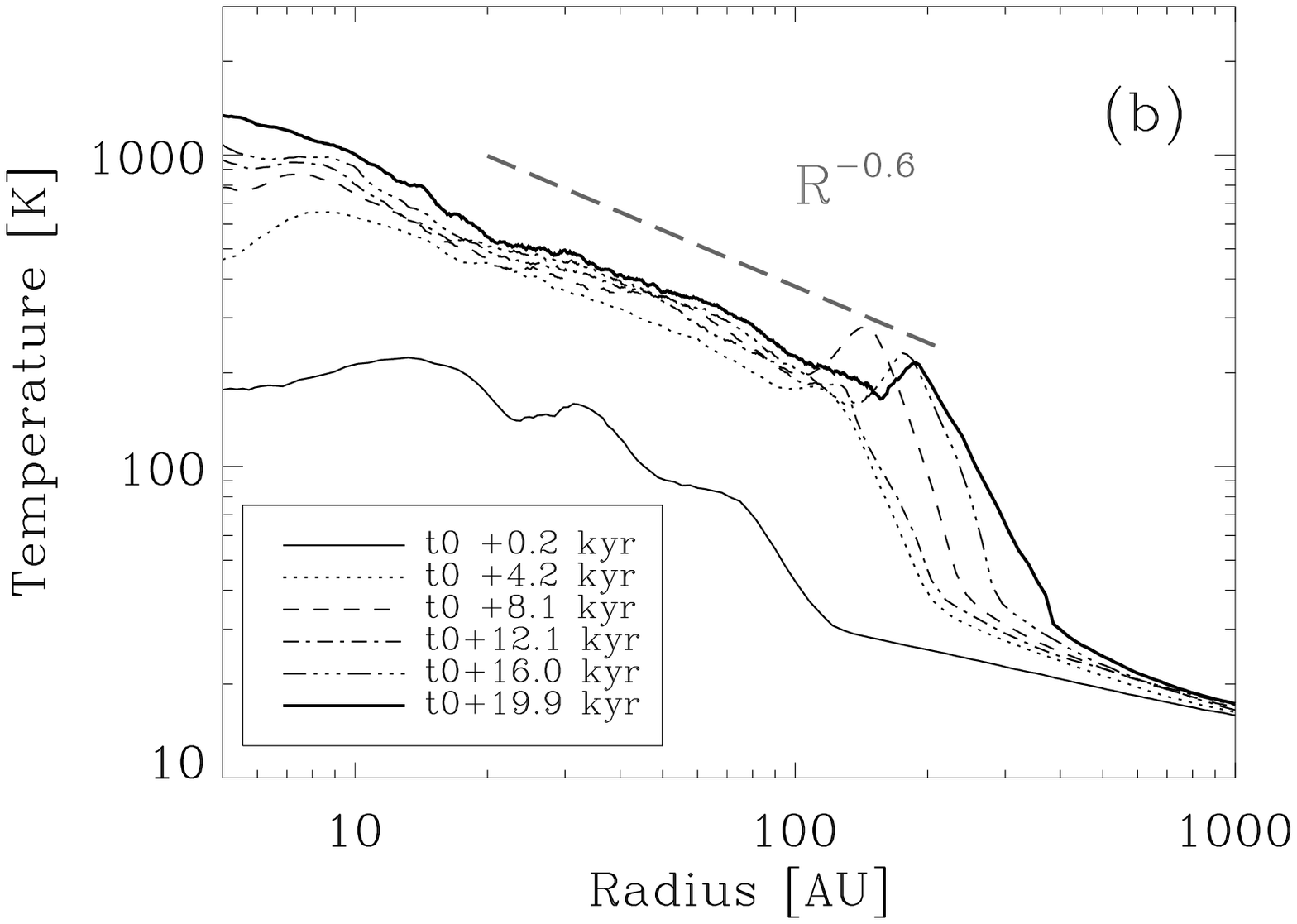, height=6cm}
\psfig{figure=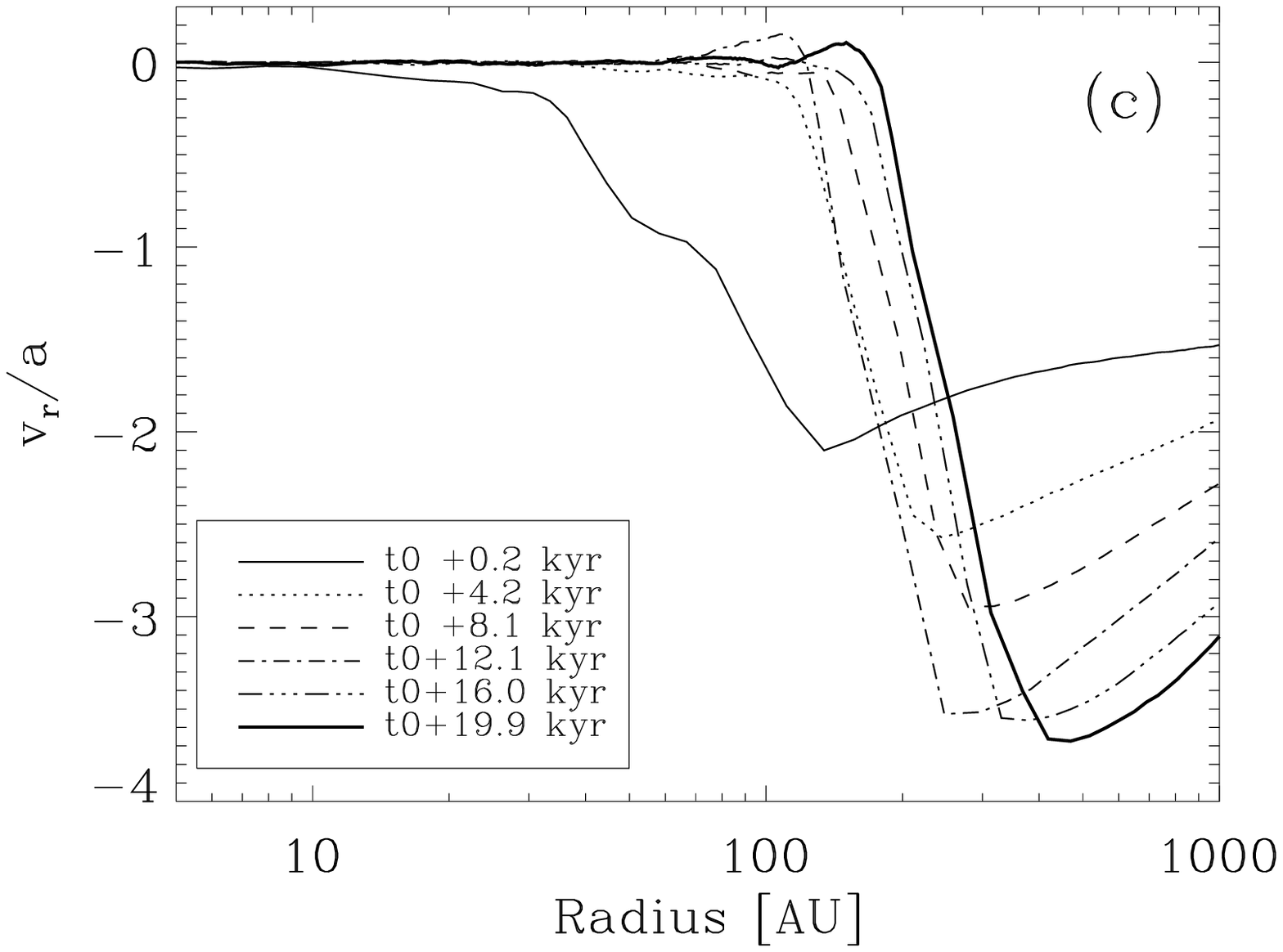, height=6cm}\\
\psfig{figure=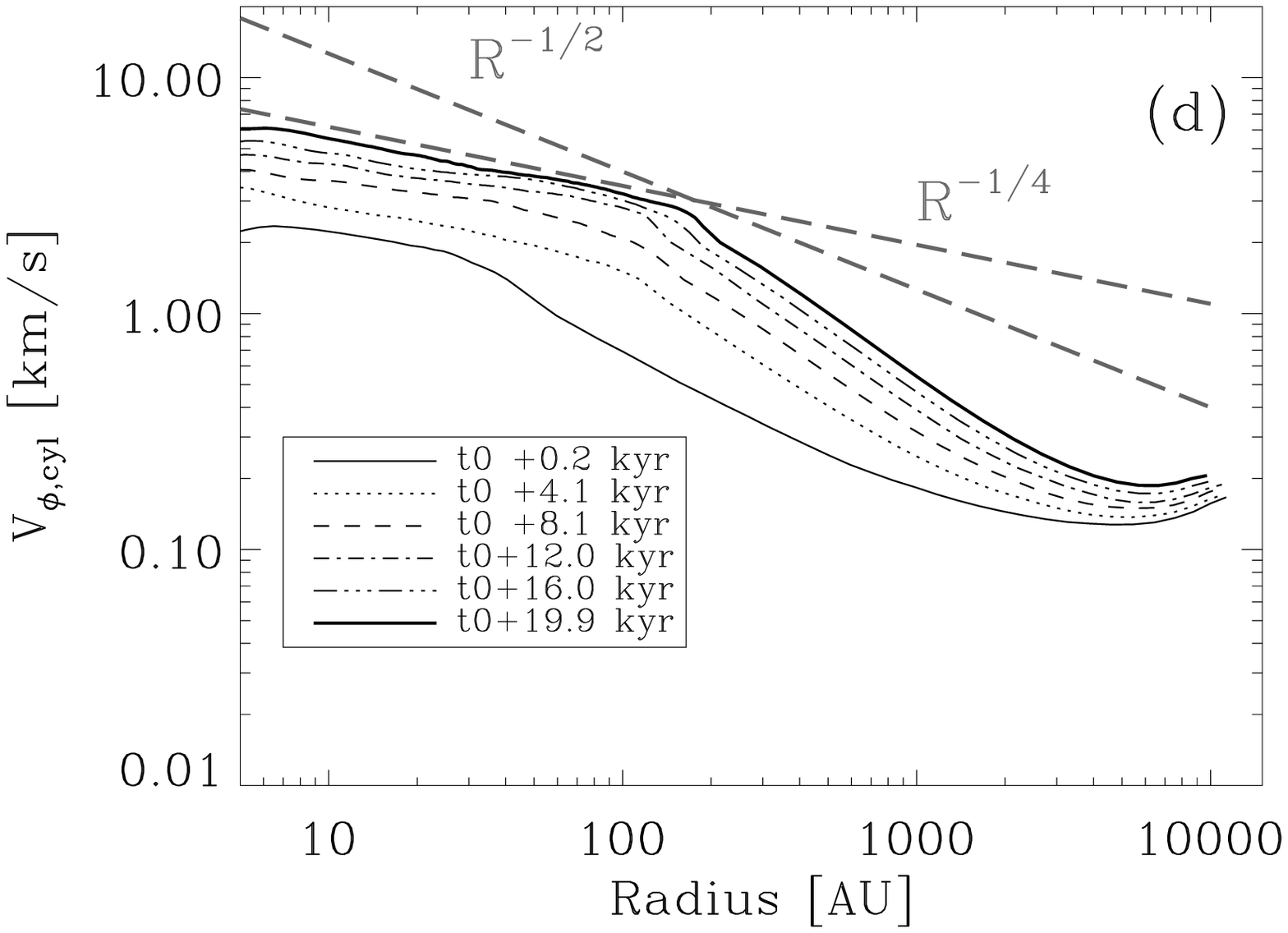, height=6cm}
\psfig{figure=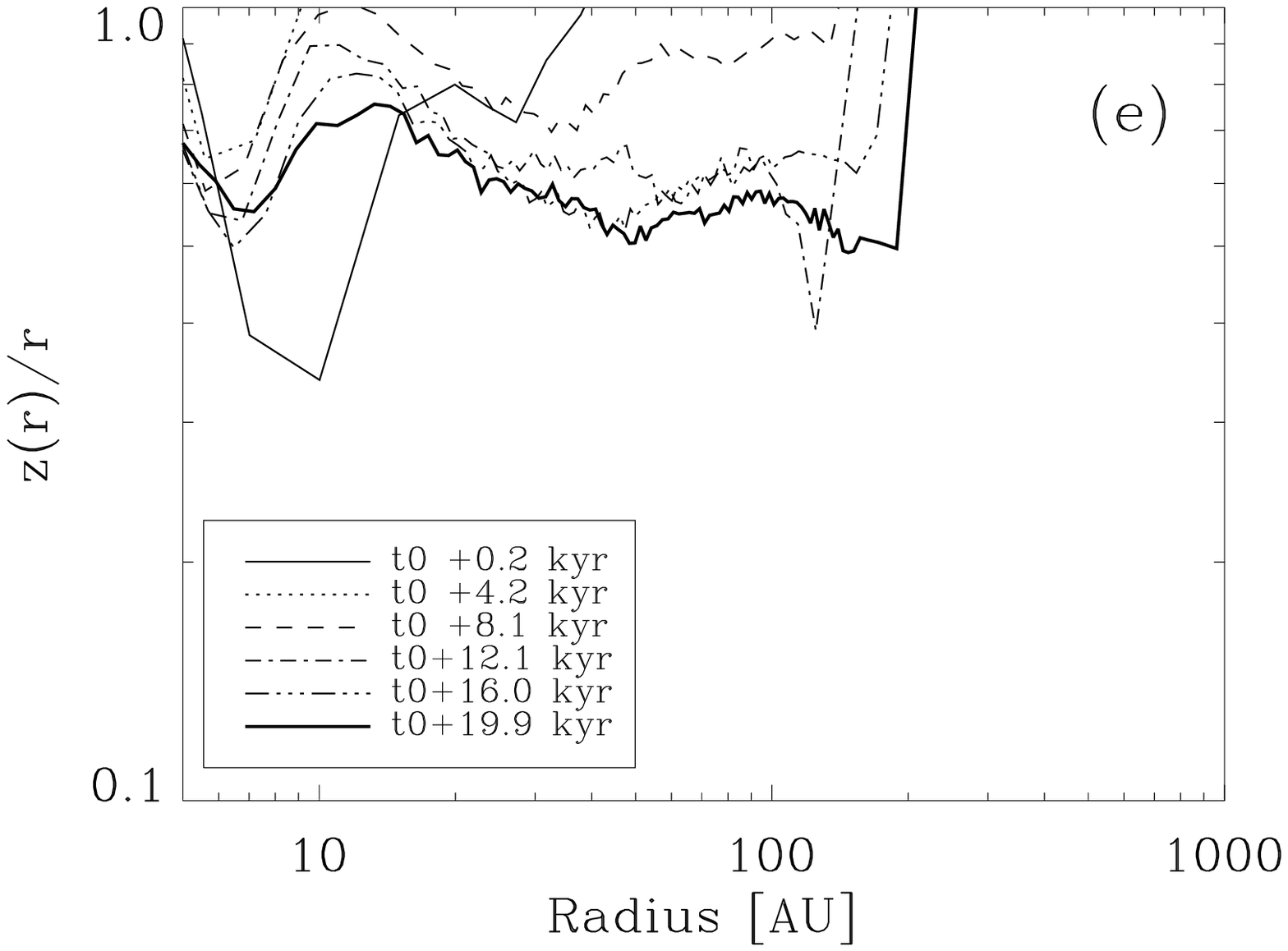, height=6cm}
\label{FIG:RUN1PROFILES}
\parbox[r]{9cm}{
\caption{Profiles from Run 1 at times $t_{_{\rm O}}+0.2\,{\rm kyr}$ (i.e. just after the primary protostar appears), $t_{_{\rm O}}+4.2\,{\rm kyr}$, $t_{_{\rm O}}+8.1\,{\rm kyr}$ (i.e. the stage at which the disc becomes well established), $t_{_{\rm O}}+12.1\,{\rm kyr}$, $t_{_{\rm O}}+16.0\,{\rm kyr}$, and $t_{_{\rm O}}+19.9\,{\rm kyr}$ (the end of the simulation). (a) The radial density profile on the equatorial plane, $\rho(r,z\!=\!0)$. The axes are logarithmic, and a line with slope $-\,2.5$ is shown to mimic the run of density in the disc between $10\;{\rm and}\;200\,{\rm AU}$. The key gives the line-styles used to represent the different times. (b) The radial temperature profile on the equatorial plane, $T(r,z\!=\!0)$. The axes are logarithmic, and a line with slope $-\,0.6$ is shown to mimic the run of temperature in the disc between $10\;{\rm and}\;200\,{\rm AU}$. (c) The radial velocity profile on the equatorial plane, $v_r(r,z\!=\!0)$. The velocity axis is linear, and velocity is given in units of the local sound speed. The radius axis is logarithmic. (d) The azimuthal velocity profile on the equatorial plane, $v_\phi(r,z\!=\!0)$. The axes are logarithmic, and lines with slopes of $-1/2$ and $-1/4$ are included to show that in the disc $v_\phi\propto r^{-1/4}$, corresponding to non-Keplerian differential rotation. Note that this plot has a larger radial extent than the others. (e) The radial profile of the disc aspect ratio, ${\bar z}(r)/r$. Both axes are linear.}}
\end{multicols}
\end{figure*}

{\sc Radial velocity.} Fig. 3(c) shows radial velocity profiles on the equatorial plane, normalised to the local isothermal sound speed, $v_r(r,z\!=\!0)/a(r,z\!=\!0)$, from Run 1. Once the disc is well established (i.e. by $t=t_{_{\rm O}}+8.1\,{\rm kyr}$), the gas in the equatorial plane is in freefall, until it hits the accretion shock at the edge of the disc ($100\;{\rm to}\;200\,{\rm AU}$) at $\sim{\rm Mach}\,4$. The gas away from the equatorial plane hits the accretion shocks bounding the disc above and below even faster, because these shocks are even deeper in the gravitational potential well of the core. Thus the gas entering the disc is typically shock-heated to at least $100\,{\rm K}$. The radial velocity of the gas in the disc is very tiny because the effective viscosity arising from gravitational torques is small, and therefore the gas takes many orbits to migrate inwards.

{\sc Azimuthal velocity.} Fig. 3(d) shows azimuthal velocity profiles on the equatorial plane, $v_\phi(r,z\!=\!0)$, from Run 1. Throughout the disc the gas is in approximate centrifugal balance; because the disc is massive and self-gravitating, this results in a profile with $v_\phi(r,z\!=\!0)\!\propto\! r^{-1/4}$ (significantly flatter than a purely Keplerian $v_\phi(r,z\!=\!0)\!\propto\! \left.r^{-1/2}\right)$. Outside the disc, the infalling matter changes from its initial solid-body $\,v_\phi(r,z\!=\!0)\!\propto\! r\,$ to $\,v_\phi(r,z\!=\!0)\!\propto\! r^{-1}$, as it falls inside $\sim 6,000\,{\rm AU}$.

{\sc Scale-height.} At each radius $r$ in the disc, we can define the vertical scale-height as
\begin{eqnarray}
{\bar z}(r)&=&\frac{\bar{\Sigma}(r)}{\bar{\rho}(r,z\!=\!0)}\,,
\end{eqnarray}
where $\Sigma(r)$ is the azimuthally averaged surface-density (as seen from $z=\pm\infty$) and $\rho(r,z\!=\!0)$ is the azimuthally averaged volume-density on the midplane$\;(z\!=\!0)$. Fig. 3(e) shows radial profiles of the disc aspect ratio, $\,{\bar z}(r)/r$, from Run 1. The aspect ratio is almost constant, at a value $\,{\bar z}(r)/r\sim 0.55$. Thus the disc is neither flat nor flaring; rather the iso-density surfaces are approximately conical. Since the midplane density varies as $\rho(r,z\!=\!0)\propto r^{-5/2}$ and the scale-height varies as ${\bar z}\propto r$, the surface-density of the disc satisfies $\,\Sigma(r)\propto r^{-3/2}$. [This is very similar to what is inferred for protoplanetary discs by \citet{Dullemond2001}, but our discs are much younger and more massive.] It follows that the mass of the disc out to radius $r$ is roughly $\,M_{_{\rm DISC}}(r)\propto r^{1/2}$, and hence the equilibrium azimuthal velocity should approximate to
\begin{eqnarray}
v_\phi(r)&\simeq&\left(\frac{G\,M_{_{\rm DISC}}(r)}{r}\right)^{1/2}\;\,\propto\;\,r^{-1/4}\,,
\end{eqnarray}
as indeed it does (see Fig. 3(d)).

\begin{figure}
\label{FIG:GROWTH}
\psfig{figure=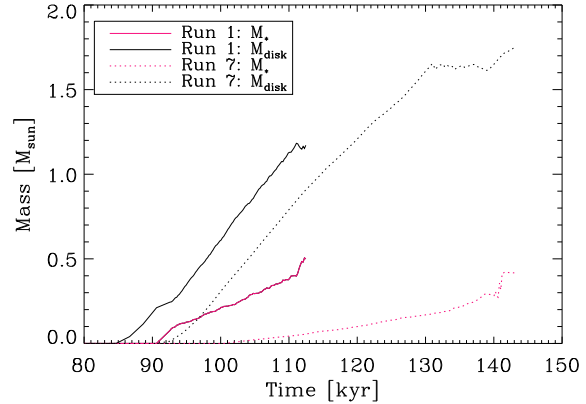, width=8.4cm}
\caption{The masses of the primary protostar, $M_\star(t)$ (magenta) and the protostellar disc, $M_{_{\rm D}}(t)$ (black), as a function of time, for Runs 1 and 7.}
\end{figure}

{\sc Disc growth.} Fig. 4 shows the growth of the primary protostar, $M_\star(t)$, and the growth of the disc, $M_{_{\rm D}}(t)$  The growth rates are roughly constant at $\dot{M}_\star\sim 2.2\times 10^{-5}\,{\rm M}_\odot\,{\rm yr}^{-1}$ and $\dot{M}_{_{\rm D}}\sim 5.0\times 10^{-5}\,{\rm M}_\odot\,{\rm yr}^{-1}$; hence the disc grows faster than the primary protostar. The disc is marginally Toomre unstable \citep[][see Fig. 7]{1964Toomre}, to the extent that weak spiral arms develop and contribute to the inward transport of material through the disc and onto the primary protostar. However, the disc is not self-regulating, in the sense that the rate of transport through the disc does not adjust so that $\dot{M}_\star\sim\dot{M}_{_{\rm D}}$. The disc is quite concentrated: typically more than $60\%$ of the disc mass has density $\rho >10^{-13}\,{\rm g}\,{\rm cm}^{-3}$. Radiative cooling times in the disc are sufficiently long that the discs undergoes gravo-acoustic oscillations in the vertical direction, in response to the continuing infall of matter.
\begin{figure}
  \epsfig{figure=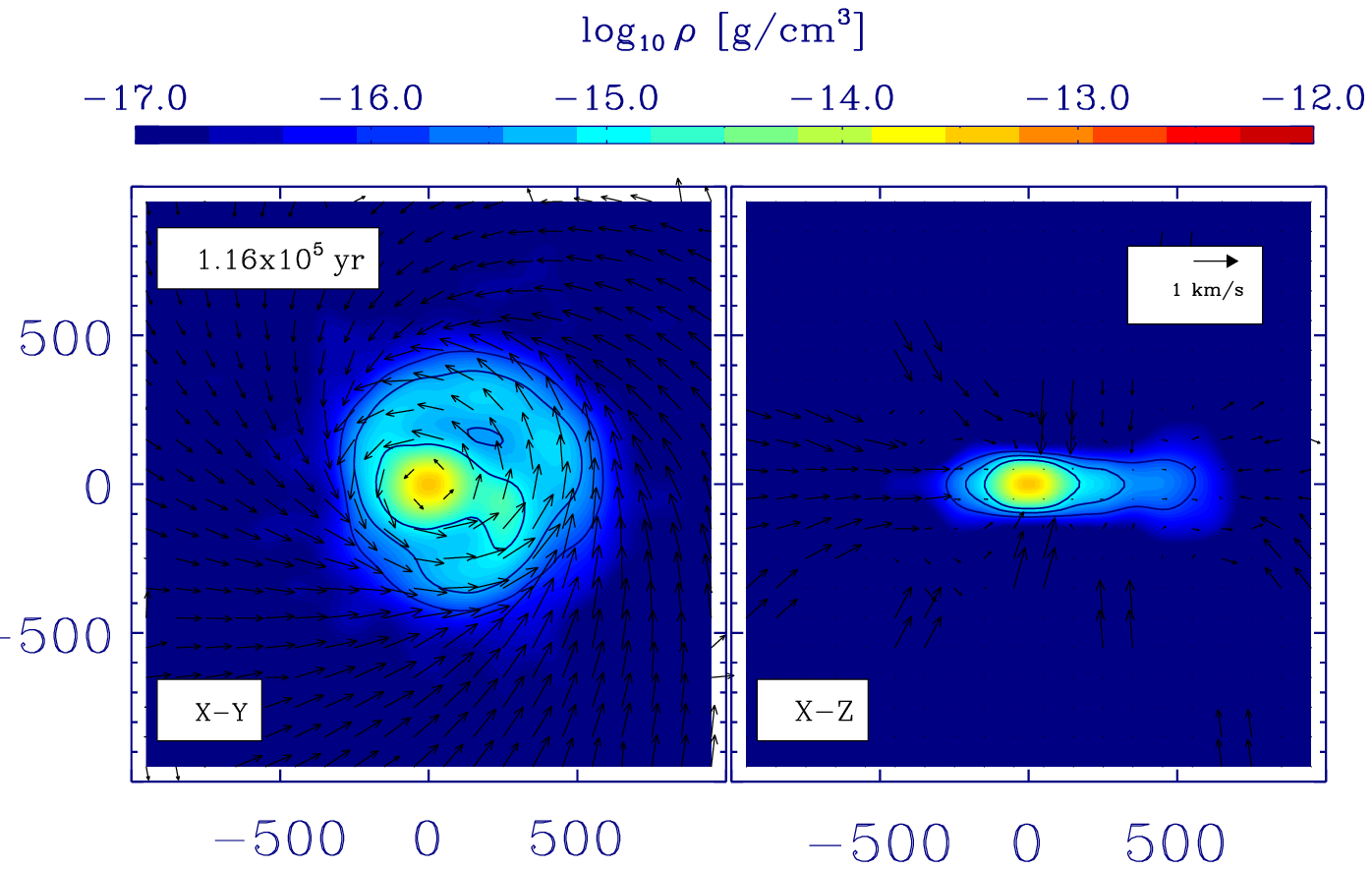, width=7.3cm}
  \epsfig{figure=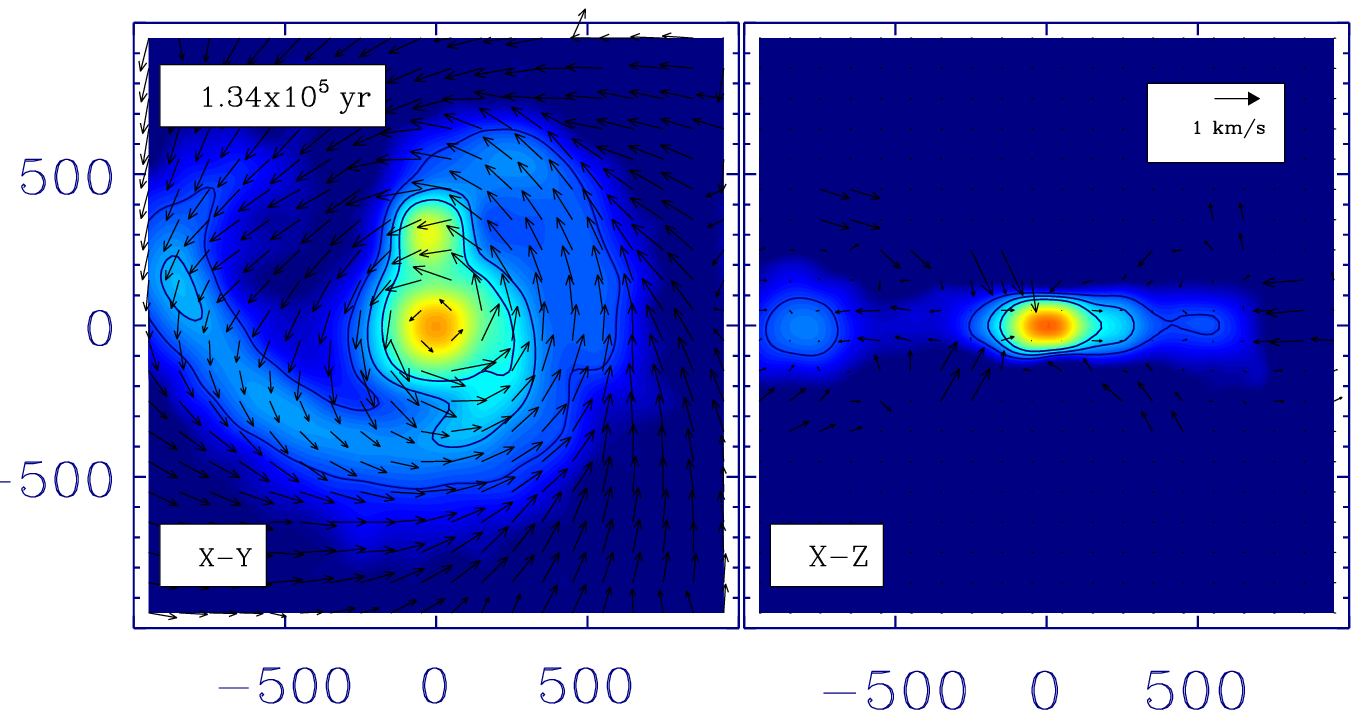,width=7.3cm}
  \epsfig{figure=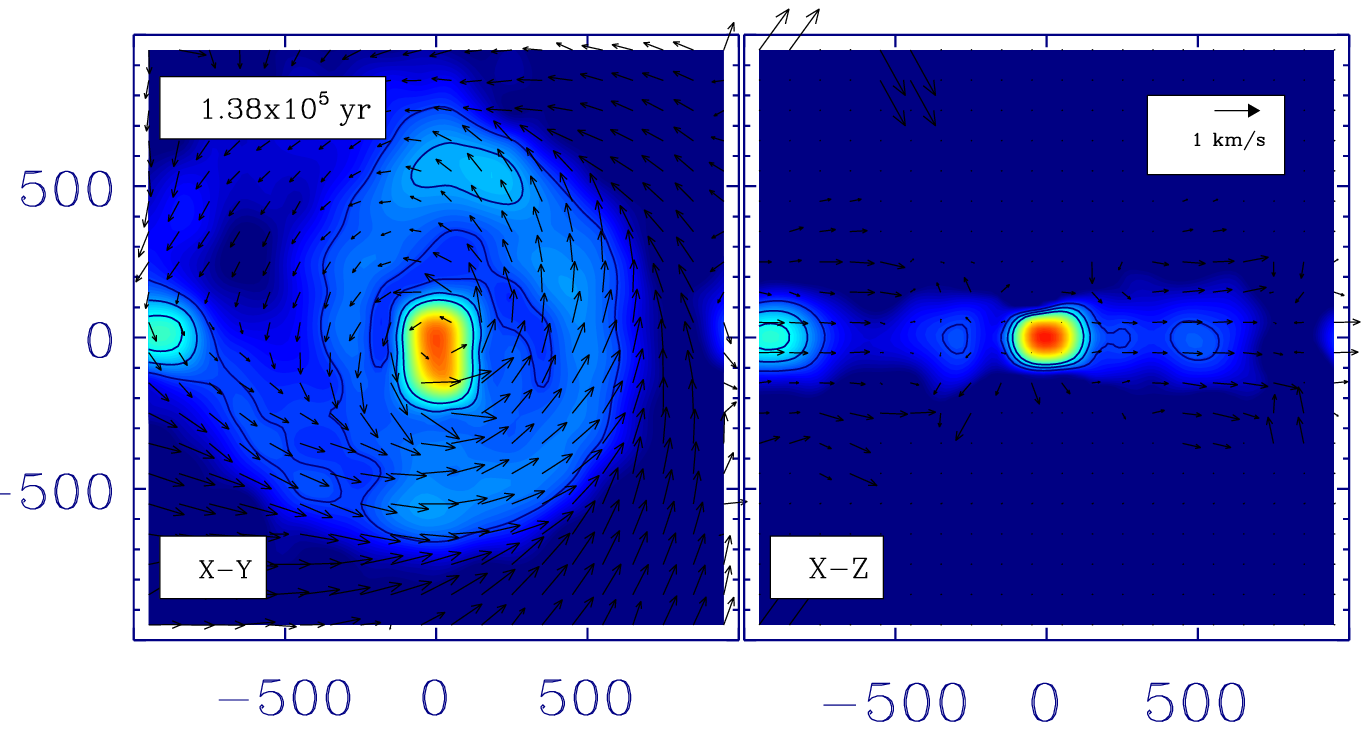,width=7.3cm}
  \epsfig{figure=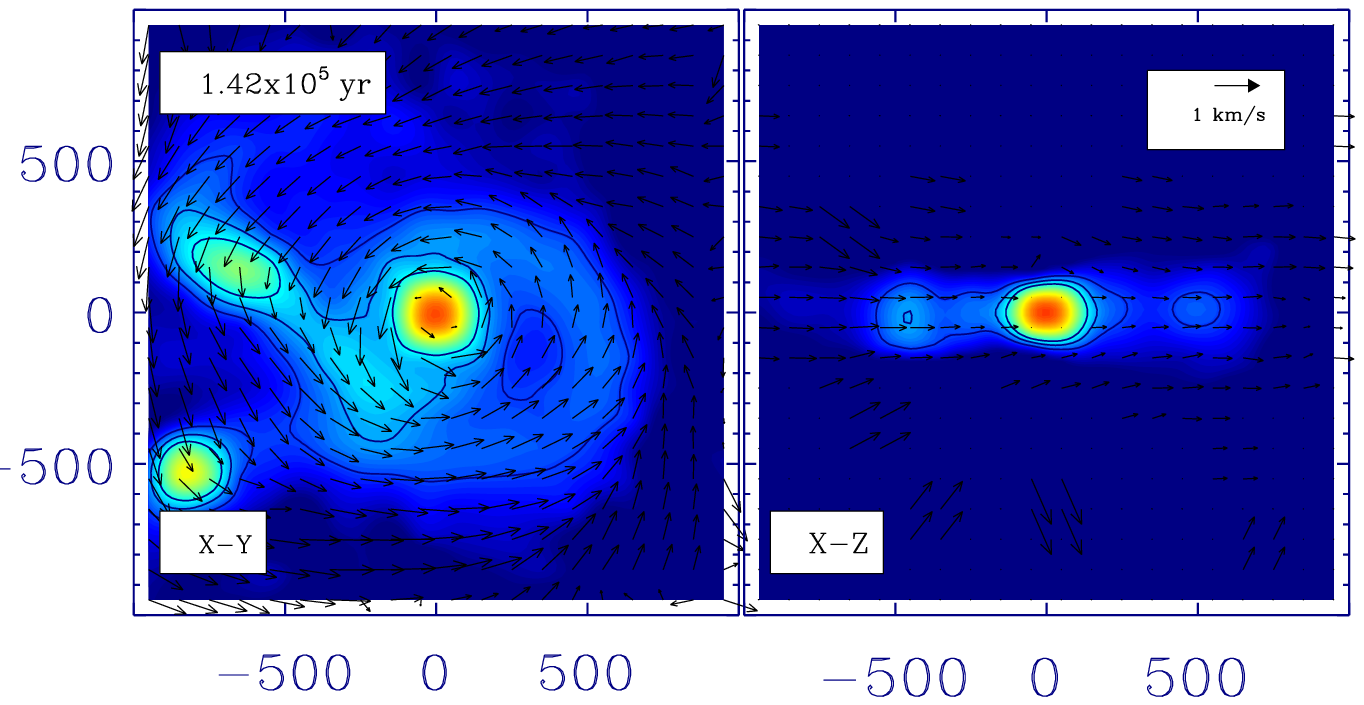,width=7.3cm}
  \epsfig{figure=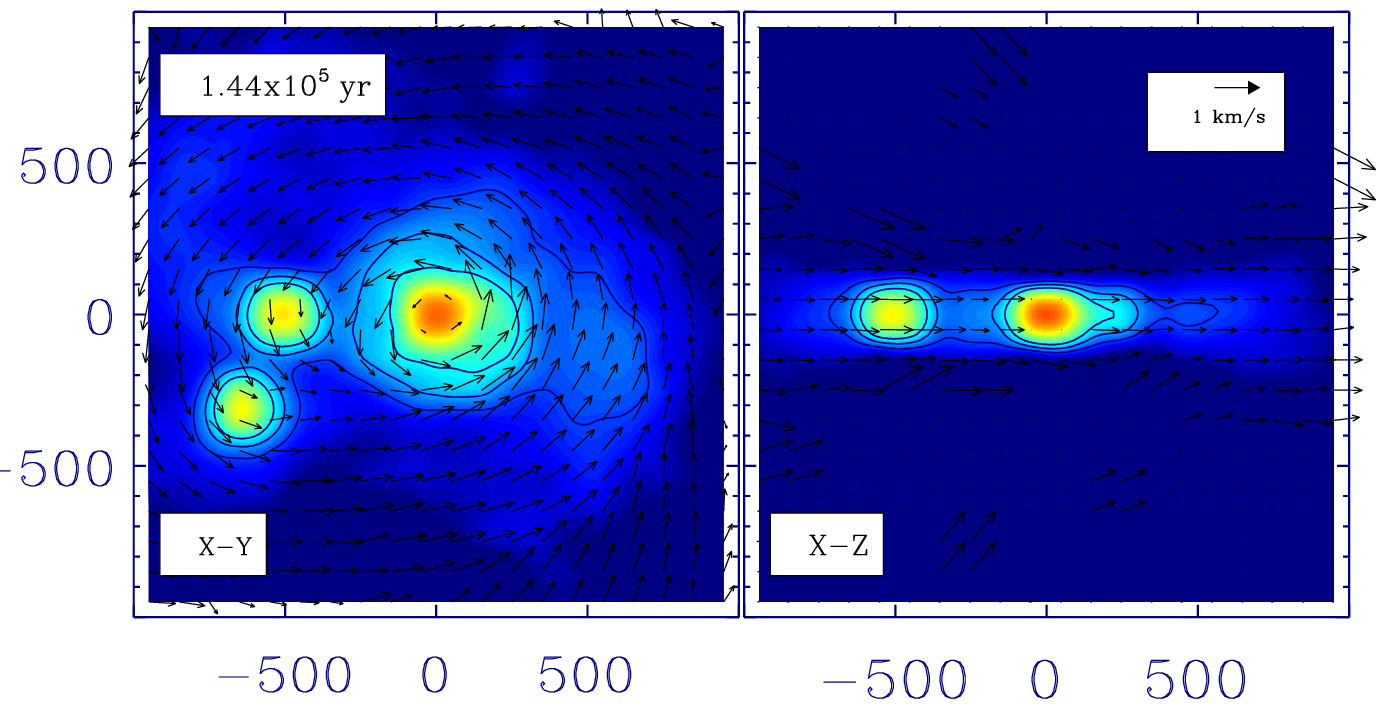,width=7.3cm}
\caption{The density on slices through the centre of the core during Run 7. The left-hand (respectively right-hand) column shows slices at $z\!=\!0\;\,\left({\rm respectively\;}y\!=\!0\right)$. The times shown are (a) $t_{_{\rm O}}+11.8\,{\rm kyr}$; (b) $t_{_{\rm O}}+29.7\,{\rm kyr}$; (c) $t_{_{\rm O}}+33.7\,{\rm kyr}$; (d) $t_{_{\rm O}}+33.7\,{\rm kyr}$, zoomed to show the first long-lived fragment; (e) $t_{_{\rm O}}+38.0\,{\rm kyr}$; and (f) $t_{_{\rm O}}+40.0\,{\rm kyr}$. In all cases the slices are $2000\,{\rm AU}\times 2000\,{\rm AU}$. The density scale ranges from $10^{-17}\;{\rm to}\;10^{-12}\,{\rm g}\,{\rm cm}^{-3}$, and contours are plotted at $10^{-16},\,3\times 10^{-16},\;{\rm and}\;10^{-15}\,{\rm g}\,{\rm cm}^{-3}$. Velocity vectors are superimposed, and the insets show a $1\,{\rm km}\,{\rm s}^{-1}$ vector.}\label{FIG:RUN7DVF}
\end{figure}

\section{A high angular momentum core}
In this section we describe the evolution of a high angular momentum core, having $\Omega_{_{\rm O}}=1.8\times 10^{-13}\,{\rm s}^{-1}$, and hence $j_{_{\rm O}}=3.3\times 10^{21}\,{\rm cm}^2\,{\rm s}^{-1}$ and $\beta=0.13$. This is Run 7.

{\sc The Isothermal Collapse phase.} The Isothermal Collapse Phase of Run 7 proceeds in the same way as for Run 1, except that, due to the higher angular momentum, departures from spherical symmetry occur sooner, and are more extensive. As a result, the Isothermal Collapse Phase ends with the formation of a protostellar disc, and the primary protostar does not form for another $15\,{\rm kyr}$, by which stage the disc extends to $700\,{\rm AU}$.

{\sc The Protostellar Phase.} Fig. 5 shows velocity vectors superimposed on false-colour images of the density field on $2000\,{\rm AU}\times 2000\,{\rm AU}$ slices through the centre of the core during Run 7. These images correspond to times from just before the formation of the primary protostar at $t=t_{_{\rm O}}$ to $t=t_{_{28}}$. They show the growth and fragmentation of the protostellar disc. Densities and velocities are computed as in Fig. 2. The features that distinguish this high-$\Omega_{_{\rm O}}$ run (Run 7) from the low-$\Omega_{_{\rm O}}$ one discussed in the preceding section (Run 1) are (a) that the primary protostar grows much more slowly and acquires most of its mass via the disc (rather than by direct infall), and (b) that the outer parts of the disc are able to fragment creating secondary protostars.

\begin{figure*}
\begin{multicols}{2}
\psfig{figure=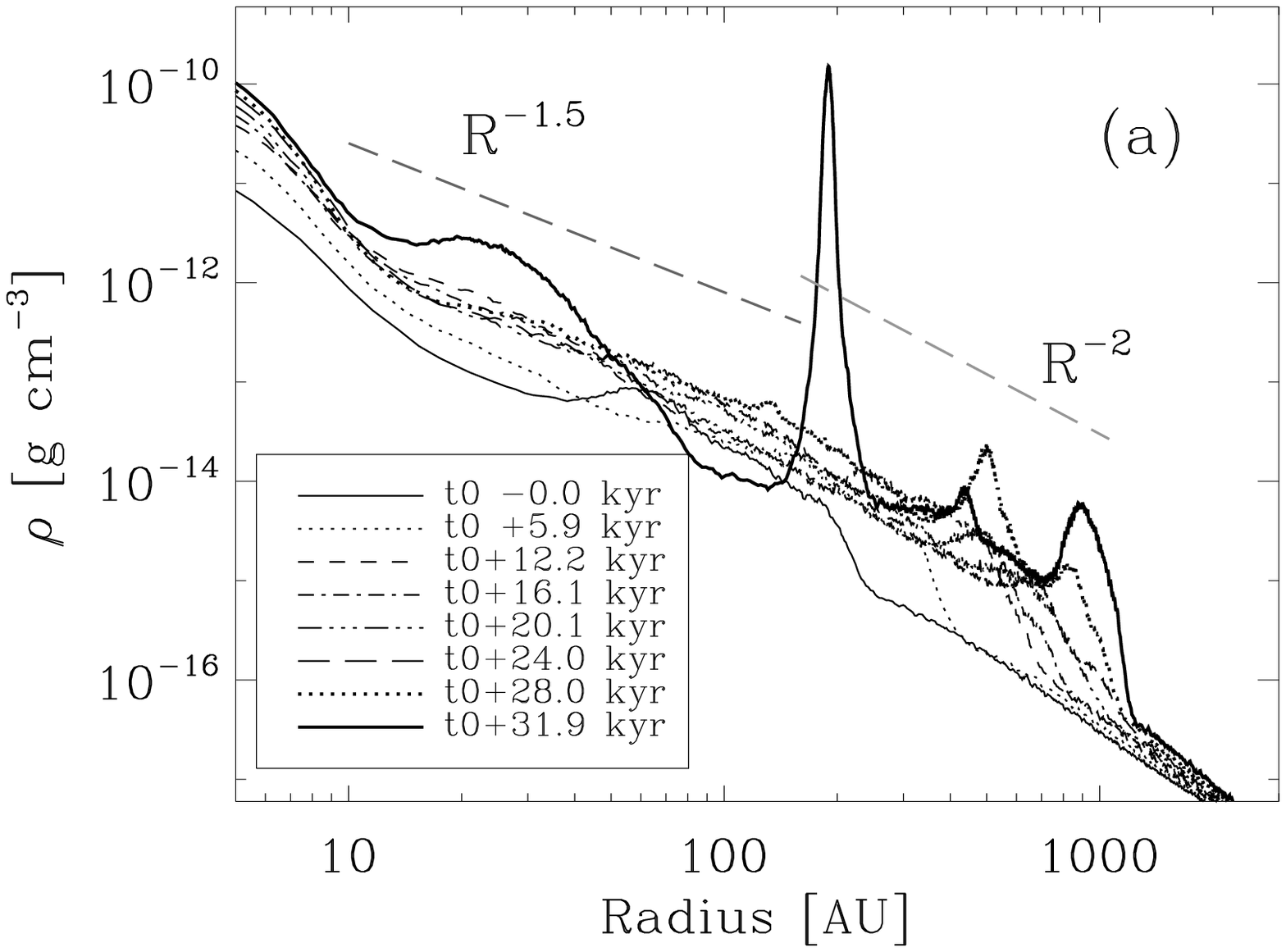, height=6cm}
\psfig{figure=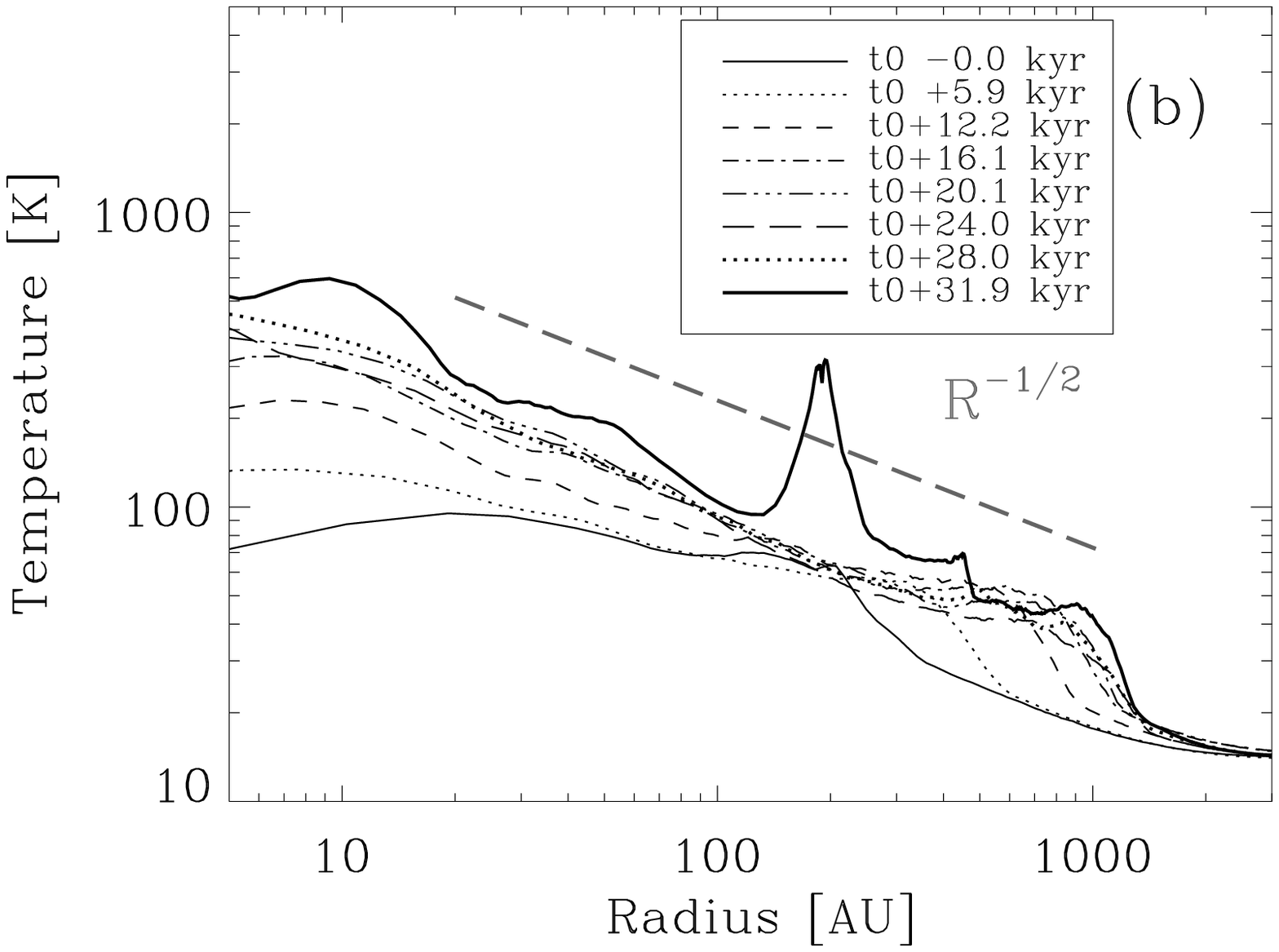, height=6cm}
\psfig{figure=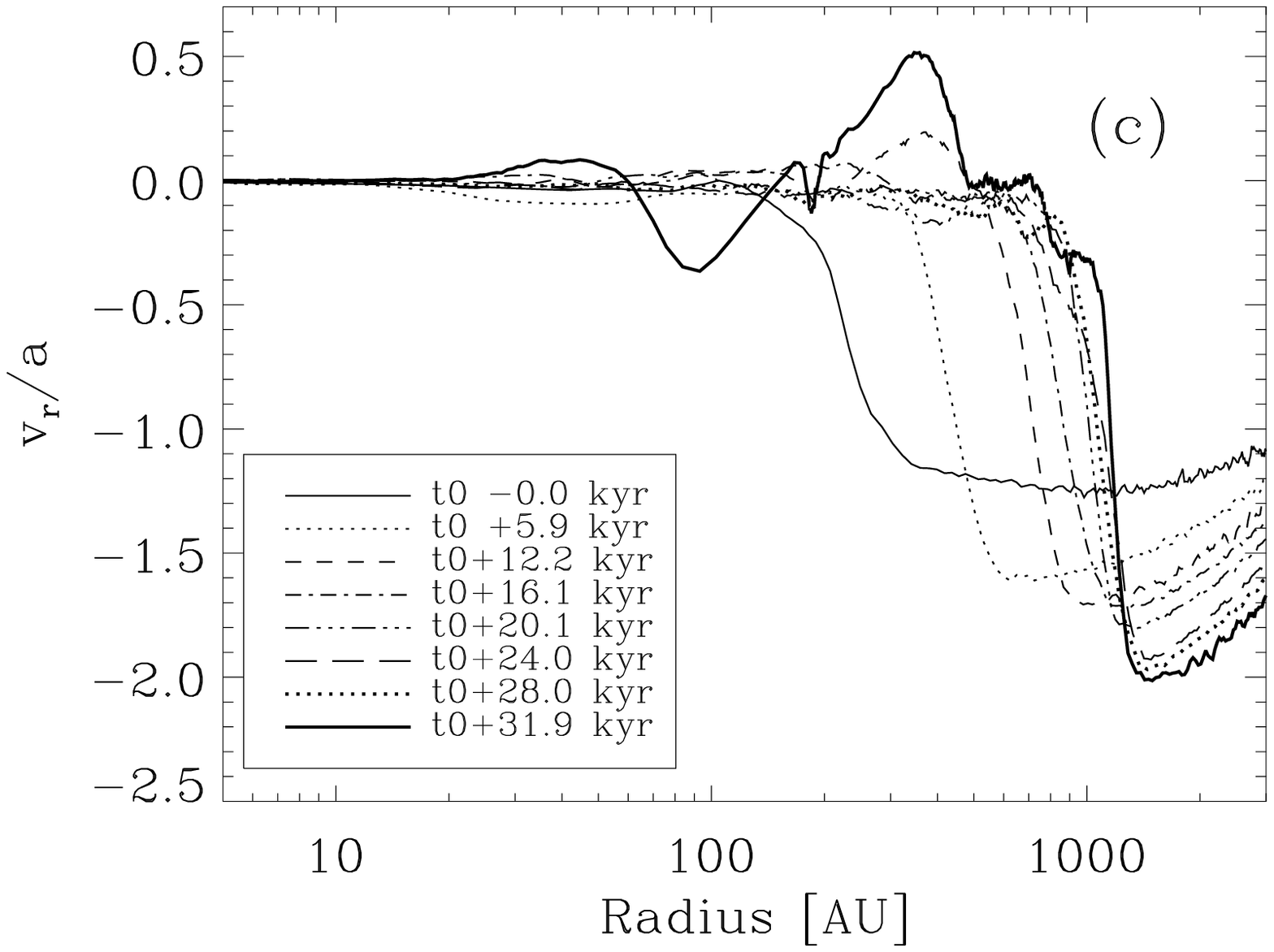, height=6cm}\\
\psfig{figure=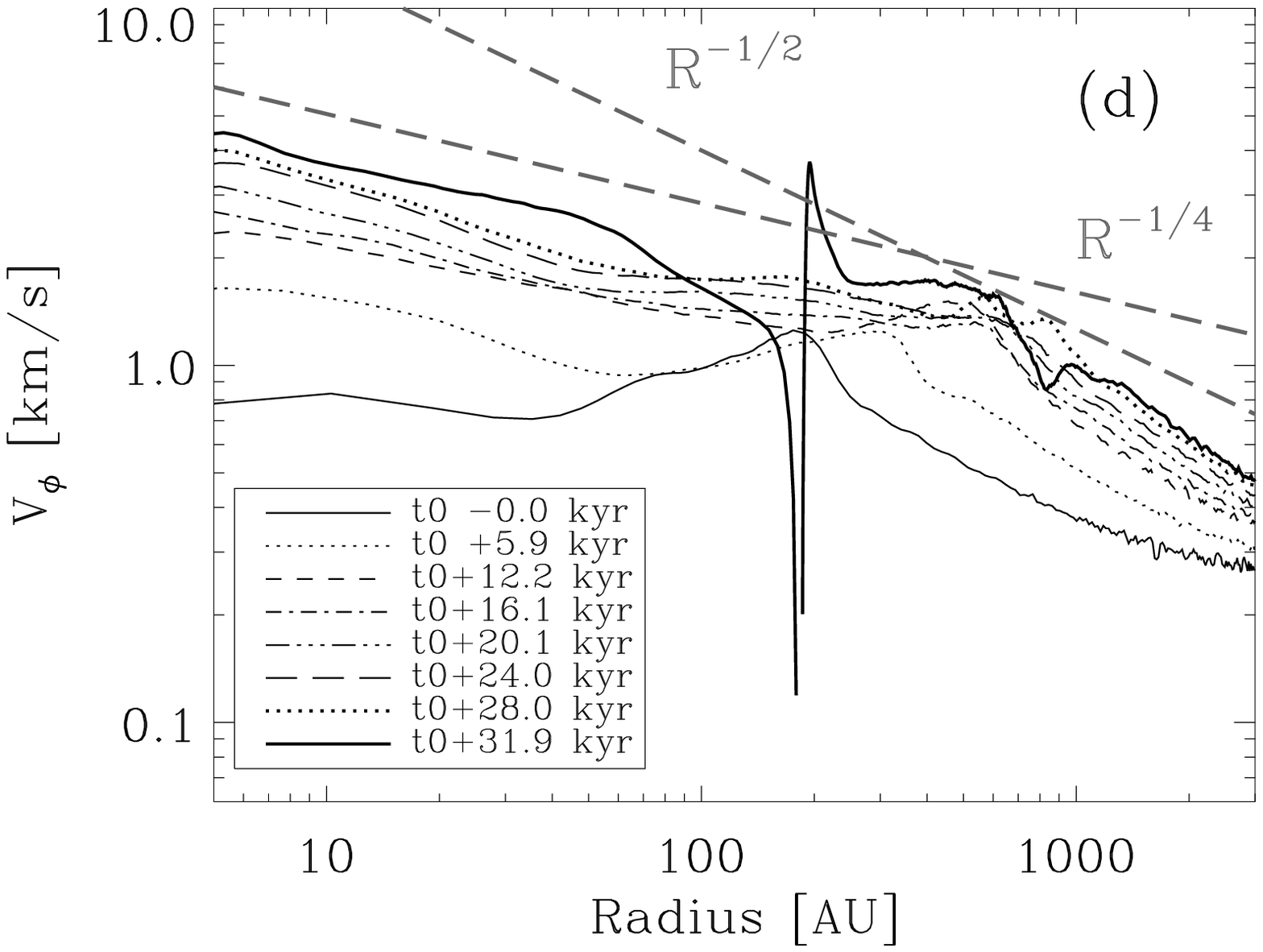, height=6cm}
\psfig{figure=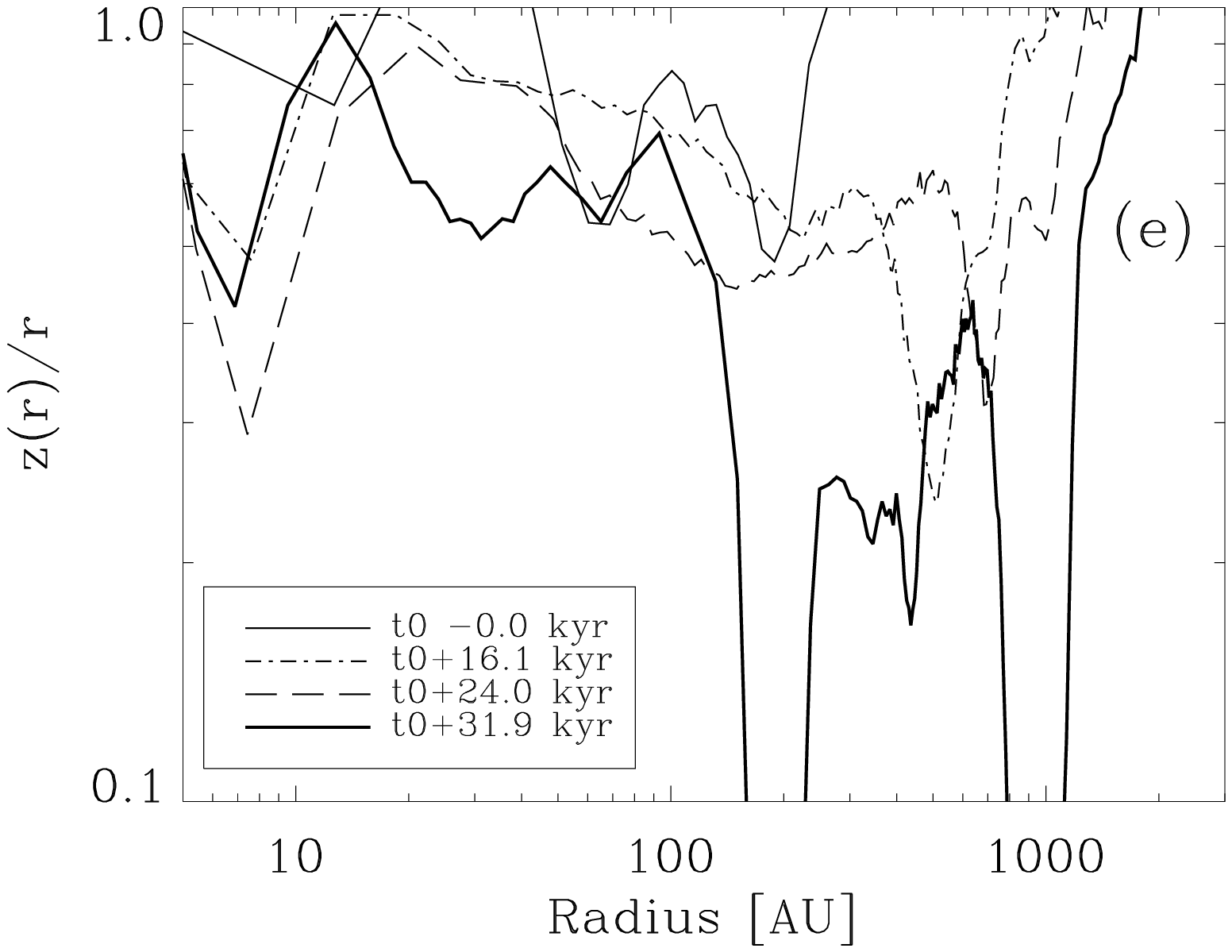, height=6cm}
\parbox[h]{9cm}{
\caption{Profiles from Run 7 at times $t_{_{\rm O}}$ (i.e. the moment at which the primary protostar forms), $t_{_{\rm O}}+5.9\,{\rm kyr}$, $t_{_{\rm O}}+12.2\,{\rm kyr}$, $t_{_{\rm O}}+16.1\,{\rm kyr}$, $t_{_{\rm O}}+20.1\,{\rm kyr}$, $t_{_{\rm O}}+24.0\,{\rm kyr}$, $t_{_{\rm O}}+28.0\,{\rm kyr}$, and $t_{_{\rm O}}+31.9\,{\rm kyr}$ (the end of the simulation). (a) The radial density profile on the equatorial plane, $\rho(r,z\!=\!0)$. The axes are logarithmic, and lines with slopes $-\,1.5$ and $-\,2.0$ are shown to mimic the run of density in the disc from $10\;{\rm to}\;100\,{\rm AU}$, and from $100\;{\rm to}\;1000\,{\rm AU}$, respectively. The key gives the line-styles used to represent the different times. (b) The radial temperature profile on the equatorial plane, $T(r,z\!=\!0)$. The axes are logarithmic, and a line with slope $-\,0.5$ is shown to mimic the run of temperature in the disc between $10\;{\rm and}\;1000\,{\rm AU}$. (c) The radial velocity profile on the equatorial plane, $v_r(r,z\!=\!0)$. The velocity axis is linear, and velocity is given in units of the local sound speed. The radius axis is logarithmic. (d) The azimuthal velocity profile on the equatorial plane, $v_\phi(r,z\!=\!0)$. The axes are logarithmic, and lines with slopes of $-1/2$ and $-1/4$ are included to show that in the disc $v_\phi\propto r^{-1/4}$, corresponding to non-Keplerian differential rotation. (e) The radial profile of the disc aspect ratio, ${\bar z}(r)/r$. The axes are linear.}}\label{FIG:RUN7PROFILES}
\end{multicols}
\end{figure*}

{\sc Density.} Fig. 6(a) shows azimuthally averaged radial density profiles on the equatorial plane, $\rho(r,z\!=\!0)$ from Run 7. Most of the mass interior to $10\,{\rm AU}$ is in the primary protostar. Outside this the disc extends to $\sim 100\,{\rm AU}$ with $\rho\propto r^{-1.5}$, and then from $\sim 100\,{\rm AU}$ to $\sim 1000\,{\rm AU}$ with $\rho\propto r^{-2.0}$. The disc in Run 7 is far more extended than that in Run 1, and its density profile is significantly shallower. The spikes superimposed on the density profiles are due to spiral arms and fragments condensing out of the disc.

{\sc Temperature.} Fig. 6(b) shows radial temperature profiles on the equatorial plane, $T(r,z\!=\!0)$, from Run 7. Throughout most of the disc, the run of temperature on the midplane approximates to $T\propto r^{-0.5}$, which is very similar to Run 1. However, in Run 7 the temperatures are lower. There are two reasons for this. First, because the disc is more extended, the material hitting the accretion shock at the boundary of the disc has fallen less far into the gravitational potential well of the core and is therefore moving more slowly; it is therefore not heated so much in the accretion shock. Second, again because the disc is more extended and therefore has lower surface density, but also because it is cooler, its cooling radiation can escape more readily. This in turn is the main reason why the disc in Run 7 is more prone to fragmentation. Since the rate of infall onto the disc greatly exceeds the rate of accretion onto the primary protostar (see below), the heating at the accretion shock again dominates the energy budget of the disc. Further heating by compression and viscous dissipation ensures that the vertical temperature gradient is negative, $dT/d|z|<0$. However, because the outer disc is very cool, we only expect to see the $10\,\mu{\rm m}$ feature in absorption from the inner disc. The spikes superimposed on the temperature profiles are due to fragments condensing out of the disc and being heated by compression.

{\sc Radial velocity.} Fig. 6(c) shows radial velocity profiles on the equatorial plane, normalised to the local isothermal sound speed, $v_r(r,z\!=\!0)/a(r,z\!=\!0)$, from Run 7. The gas in the equatorial plane is close to freefall until it hits the accretion shock at the outer edge of the disc. Because the disc is so extended, the material hitting the accretion shock is not moving very fast; for example, at the outer edge of the disc it is moving with $v_r\,<{\rm Mach}\,2$. Thus the gas entering the disc is only heated mildly. The radial velocity of the gas in the disc is small, although there are significant perturbations in the vicinities of fragments.

{\sc Azimuthal velocity.} Fig 6(d) shows azimuthal velocity profiles on the equatorial plane, $v_\phi(r,z\!=\!0)$, from Run 7. Throughout the disc the gas is in approximate centrifugal balance, but because the gravitational field is dominated by the mass of the disc, rather than the mass of the primary protostar, the profile is significantly flatter than Keplerian, i.e. $0\la (-d\ln v_\phi/d\ln r)\la 0.25$, except for where it is perturbed by a forming fragment. Outside the disc, the infalling matter changes from its initial solid-body $\,v_\phi(r,z\!=\!0)\!\propto\! r\,$ to $\,v_\phi(r,z\!=\!0)\!\propto\! r^{-1}$, as it falls inside $\sim 10,000\,{\rm AU}$.

{\sc Scale-height.} Fig. 6(e) shows radial profiles of the disc aspect ratio, $\,{\bar z}(r)/r$, from Run 7. The aspect ratio is almost constant at a value $\,{\bar z}(r)/r\sim 0.6$ within the inner disc ($r<200{\rm AU}$), and it falls off slightly at larger radii, where it is strongly perturbed by spiral arms and fragments. Again it is neither flat nor flaring; the iso-density surfaces are approximately conical. Since the midplane density follows $1.5\la (-d\ln \rho/d\ln r)\la 2.0$ and the scale-height varies as ${\bar z}\propto r$, the surface-density of the disc follows $0.5\la (-d\ln \Sigma/d\ln r)\la 1.0$, and the disc mass follows $0.5\la (d\ln M_{_{\rm DISC}}/d\ln r)\la 1.0$. The azimuthal velocity should approximate to $v_\phi\sim\left(GM_{_{\rm DISC}}(r)/r\right)^{1/2}$, hence $0\la (-d\ln v_\phi/d\ln r)\la 0.25$, as indeed it does (see Fig. 6(d)), modulo the fluctuations due to forming fragments.

{\sc Disc growth.} Fig. 4 shows the growth of the primary protostar, $M_\star(t)$, and the growth of the disc, $M_{_{\rm D}}(t)$, in Run 7. Until the disc starts to fragment, and then a fragment merges with the primary protostar, the growth rates are roughly constant at $\dot{M}_\star\sim 0.7\times 10^{-5}\,{\rm M}_\odot\,{\rm yr}^{-1}$ and $\dot{M}_{_{\rm D}}\sim4.7\times 10^{-5}\,{\rm M}_\odot\,{\rm yr}^{-1}$. Thus the disc not only forms before the primary protostar, but it also grows much faster.

{\sc Fragmentation.} Because the disc is massive, extended and cool, it becomes Toomre unstable ($Q_{_{\rm T}}\la 1$; see Fig.7), and strong spiral arms develop. Where the cooling time in the disc is shorter than the dynamical times, the disc fragments \citep{Gammie2001, Rice2003}. The disc is only weakly concentrated: typically less than $20\%$ of the disc mass has density $\rho >10^{-13}\,{\rm g}\,{\rm cm}^{-3}$, and most of this dense gas is located in the spiral arms and fragments.

{\sc Individual fragments.} The first fragment condenses out at $t\sim t_{_{\rm O}}+30\,{\rm kyr}$ at $\sim 200\,{\rm AU}$. This fragment can be seen on Fig. 5(b) and Fig. 8. At its inception this fragment has mass $\sim 0.05\,{\rm M}_\odot$. Subsequently it grows in mass, but at the same time it spirals inwards and at $t\sim t_{_{\rm O}}+38\,{\rm kyr}$ it merges with the primary protostar. Two further fragments condense out at $t\sim t_{_{\rm O}}+40\,{\rm kyr}$, at $\sim 400\,{\rm AU}$ and $\sim 600\,{\rm AU}$ (see fourth and fith panel of Fig. 5). The simulation is terminated soon after their formation, so it is unclear whether they also merge with the primary protostar, or whether they survive as secondary companions.
\begin{figure}
  \psfig{figure=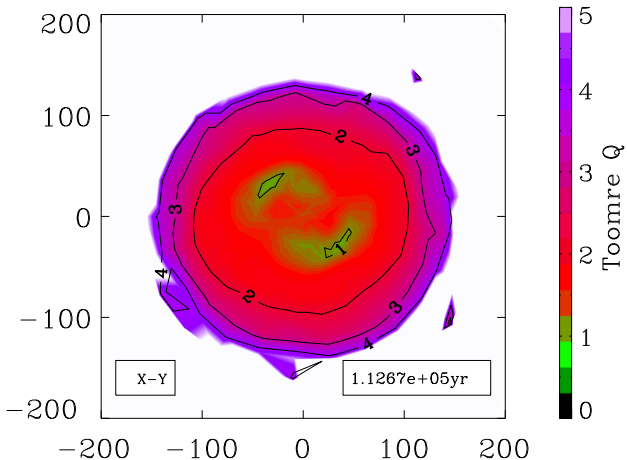, width=8cm}
  \psfig{figure=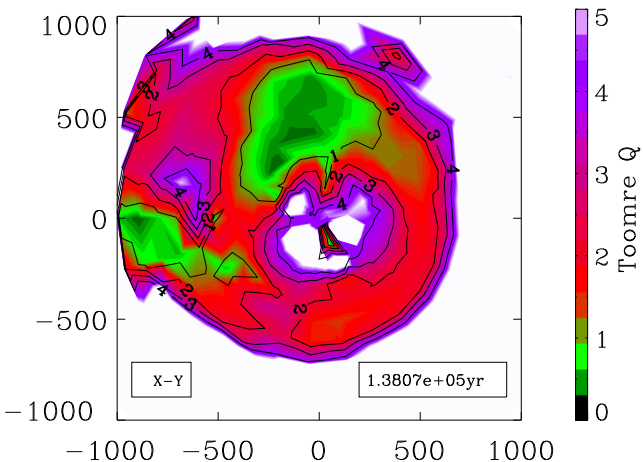,width=8cm}
  \caption{The Toomre parameter, $Q_{_{\rm T}}$, calculated for the face-
    on projections of Run 1 at $t_0+18.3$kyr (left plot), and Run 7 at $t_0
    +21.5$kyr (right plot).}\label{FIG:TOOMRE}
\end{figure}

{\sc Circum-fragmentary discs.} Fig. 8 shows velocity vectors superimposed on false-colour images of the density field in $140\,{\rm AU}\times 140\,{\rm AU}$ slices through the first fragment to form. Densities and velocities are computed as in Fig. 2, but using co-ordinates measured relative to the centre of mass of the fragment. We see that the fragment is surrounded by a well-defined, rotationally supported circum-fragmentary disc (CFD). Very similar discs are found around the other fragments. Typically they have radii between $\sim 40\,{\rm AU}$ and $\sim 60\,{\rm AU}$.

{\sc Transport of angular momentum.} Although most of the core mass infalls directly onto the disc, the spiral arms in the disc are effective at transporting angular momentum outwards, by means of gravitational torques, thereby allowing material to spiral inwards and accrete onto the primary protostar. Fragmentation further enhances the transport of angular momentum. This is because CFDs rotate in the same sense as the mother disc. Consequently, on the inside of a CFD (i.e. the side closest to the primary protostar) the shear between the material in the CFD and the material in the mother disc acts to slow down the material in the mother disc, thereby reducing its angular momentum. Conversely, on the outside of the CFD (i.e. further from the primary protostar) the shear between the material in the CFD and the material in the mother disc acts to speed up the material in the mother disc, thereby increasing its angular momentum.

{\sc Merging.} Efficient transport of angular momentum by the CFD may cause a fragment to spiral inwards and merge with the primary protostar. However, it is possible that the inspiral is being artificially accelerated by our imposition of adiabaticity at densities $\rho>10^{-13}\,{\rm g}\,{\rm cm}^{-3}$, since this forces a fragment and its CFD to remain hot, extended and therefore very dissipative, rather than allowing them to continue contracting towards protostellar densities.

{\sc Disc mineralogy.} Whereas in Run 7 the azimuthally averaged disc temperature never rises above $\sim 200\,{\rm K}$, in Run 1 it rises above $\sim 500\,{\rm K}$, and locally it reaches $\sim 800\,{\rm K}$, sufficiently high to produce crystalline silicates. Thus we might expect crystalline silicates to be present in discs forming from low-$\Omega_{_{\rm O}}$ cores, and hence to give rise to characteristic features in the SEDs of single protostars. Conversely, we might expect crystalline silicates to be rare in the cold, extended discs forming from high-$\Omega_{_{\rm O}}$ cores, and hence their characteristic features should be weak in -- or even absent from -- the SEDs of multiple protostars. The existence of a link between the amount of crystalline dust which can be found in protoplanetary discs and the initial conditions within the parental molecular cloud core has already been suggested by \citet{Dullemond2006}.
\begin{figure}
\label{FIG:FRAGDISC}
\psfig{figure=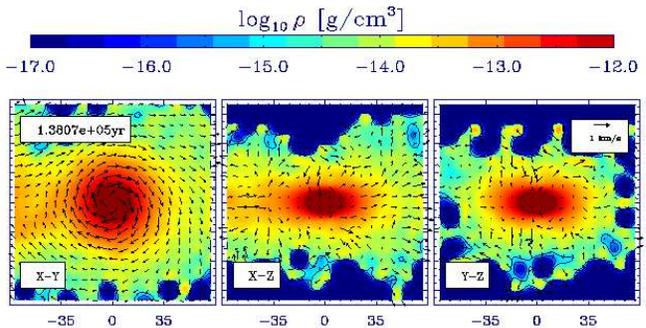, width=8.7cm}
\caption{The density field on $140\,{\rm AU}\times 140\,{\rm AU}$ slices through the centre of the first fragment to form during Run 7. The slice thickness is $6{\rm AU}$ and the time shown is $t_{_{\rm O}}+33.7\,{\rm kyr}$. The density scale ranges from $10^{-17}\;{\rm to}\;10^{-12}\,{\rm g}\,{\rm cm}^{-3}$, and contours are plotted at $10^{-16},\,3\times 10^{-16}\;{\rm and}\;10^{-15}\,{\rm g}\,{\rm cm}^{-3}$. Velocity vectors are superimposed, and the insets show a $1\,{\rm km}\,{\rm s}^{-1}$ vector.}
\end{figure}
\begin{figure}
\label{FIG:INTERMED}
\psfig{figure=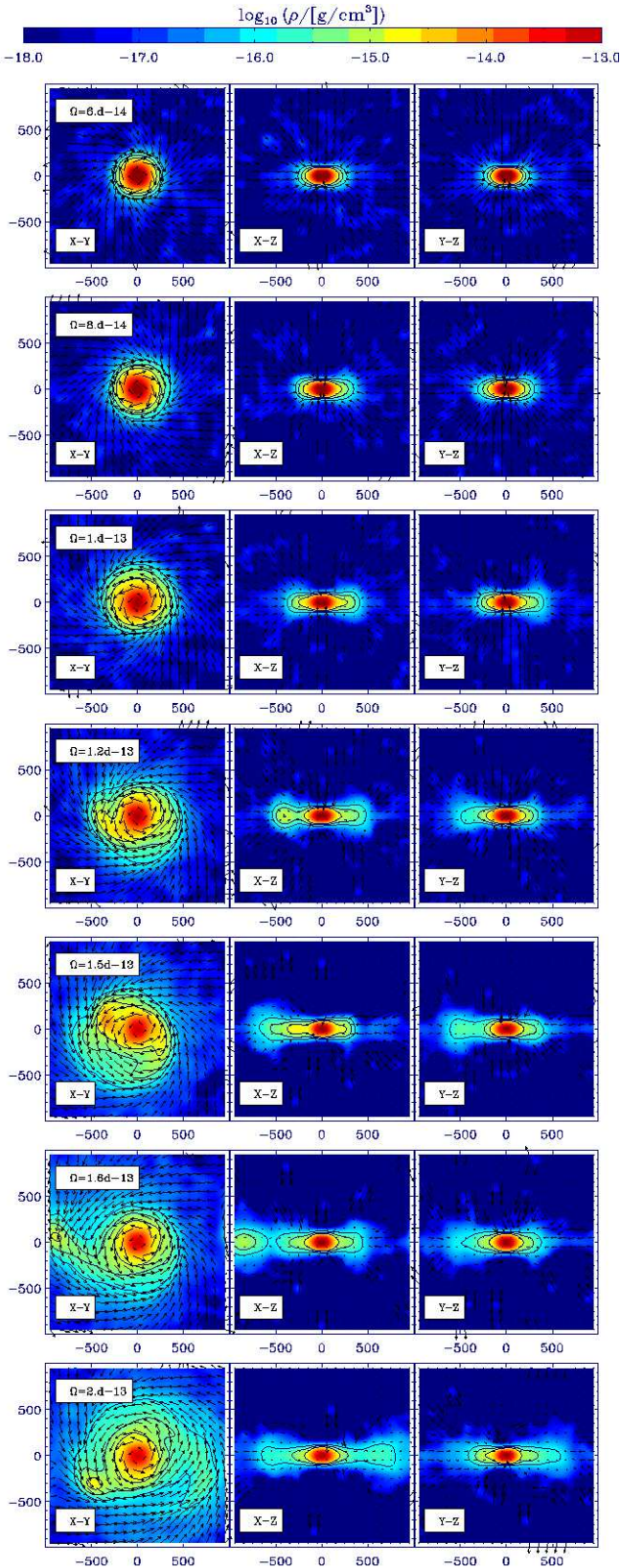, width=8.7cm}
\caption{Disc formation as a function of core angular momentum.
  For all seven runs, we show the density of the discs within slices 
  of thickness $10{\rm AU}$ about the disc midplance ($z=0$) at $t_\mathrm{final}$ 
  on the same scale. From top (Run 1) to bottom (Run 7) 
  the core rotation $\Omega_0$ and thus the total initial core
  angular momentum is increasing.}
\end{figure}

\section{Cores with intermediate angular momenta}
Referring now to the complete ensemble of Runs (1 to 7), we note that, as $\Omega_{_{\rm O}}$ increases, there is an essentially monotonic shift in the outcome; small departures from monotonicity at high values of $\Omega_{_{\rm O}}$ can be attributed to the stochastic nature of fragmentation. Key parameters are presented  in Table 1.

For low $\Omega_{_{\rm O}}$, a substantial primary protostar is formed, surrounded by a compact, thick, hot disc; the disc is unable to fragment, and simply accretes onto the primary protostar. As $\Omega_{_{\rm O}}$ is increased, the growth rate of the primary protostar decreases, not least because it acquires an increasing fraction of its mass via the disc (rather than by direct infall). At the same time the disc becomes more massive, more extended and cooler. Eventually, for $\Omega_{_{\rm O}}>1.35\times 10^{-13}\,{\rm s}^{-1}$ (Runs 5, 6 and 7), the disc becomes unstable against fragmentation, i.e. it satisfies both the Toomre and Gammie conditions \citep{1964Toomre,Gammie2001}. We are unable to ascertain whether all the resulting fragments merge with the primary protostar, or whether some of them survive as secondary companions.

As $\Omega_{_{\rm O}}$ increases, the time, $t_{_{\rm O}}$, at which the primary protostar forms increases, and its subsequent growth rate, $\dot{M}_\star$, decreases. Consequently, the overall pattern of accretion shifts from one in which the primary protostar forms first, and then the protostellar disc (low $\Omega_{_{\rm O}}$), to one in which the disc forms before the primary protostar (high $\Omega_{_{\rm O}}$).

Fig. 7 shows the Toomre parameter, $Q_{_{\rm T}}$, calculated for the face-on projections of Run 1 at $t_0+18.3$kyr (lefthand plot), and Run 7 at $t_0+21.5$kyr (right-hand plot). In the warm, compact disc from Run 1, the lowest values of $Q_{_{\rm T}}$ are in the range $1\la Q_{_{\rm T}}\la 3$, so that the disc develops weak spiral arms that transport angular momentum by exerting gravitational torques, but it does not fragment \citep[cf.][]{LinPringle1990, Laughlin1994, Lodato2004, Mejia2005}. In the cool, extended disc from Run 7, $Q_{_{\rm T}}$ falls below 1, leading to the growth of massive spiral arms and then fragmentation. {\bf This is consistent with recent findings of \citet{Rafikov2005, Matzner2005, Whitworth2006, Stamatellos2007a, Kratter2008, Rice2009, Clarke2009}, who suggest that in the inner disc ($r \la 50 {\rm AU}$) fragmentation is highly unlikely, whereas it is possible at larger radii.}

\section{Comparison with observations}\label{SEC:OBS}
\begin{figure}
\label{FIG:OBS}
\psfig{figure=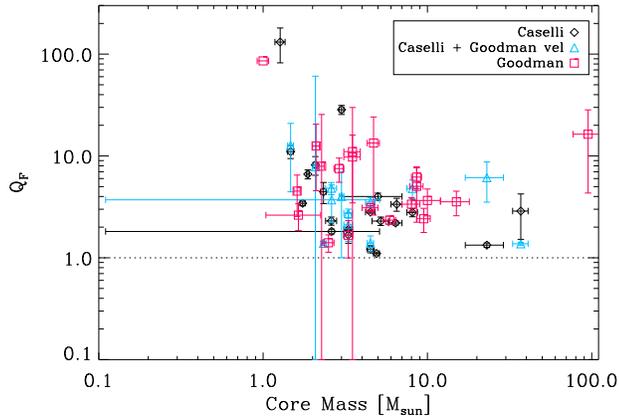, width=8.6cm}
\caption{The parameter $Q_{_{\rm F}}$ (see Eqn. \ref{EQN:MULT}) is plotted against core mass, $M_{_{\rm O}}$, for the core samples of \citet{Goodman1993}, \citet{Barranco1998}, and \citet{Caselli2002}. Cores with $Q_{_{\rm F}}<1$ are predicted to undergo disc fragmentation.}
\end{figure}

In the Appendix we develop an analytic model for the early growth and fragmentation of a protostellar disc, in terms of its mass, $M_{_{\rm O}}$, its initial radius, $R_{_{\rm O}}$, and its initial angular rotation speed, $\Omega_{_{\rm O}}$. We show that cores with $Q_{_{\rm F}}<1$, where
\begin{eqnarray}\nonumber
Q_{_{\rm F}}\!\!&\!\!=\!\!&\!\!\left(\!\frac{M_{_{\rm O}}}{6.1{\rm M}_\odot}\!\right)^{\frac{35}{92}}\!\left(\!\frac{R_{_{\rm O}}}{17,000{\rm AU}}\!\right)^{-\frac{133}{92}}\!\left(\!\frac{\Omega_{_{\rm O}}}{1.35\!\!\times\!\! 10^{-13}\,{\rm s}^{-1}\!}\right)^{-1},\\\label{EQN:MULT}
\end{eqnarray}
should be prone to disc fragmentation, whilst material is still infalling from the core envelope. We can apply this criterion to the samples of cores observed by \citet{Goodman1993}, \citet{Barranco1998}, and \citet{Caselli2002}, for which vales of $M_{_{\rm O}}$, $R_{_{\rmn O}}$ and $\Omega_{_{\rm O}}$ are listed. The resulting values of $Q_{_{\rm F}}$ are plotted against $M_{_{\rm O}}$ on Fig. 10.

Although a  few cores have $Q_{_{\rm F}}\sim 1$, none have $Q_{_{\rm F}}<1$. We infer that none of these cores is strongly predisposed to fragment whilst material is still infalling from the core envelope. The observed angular speeds are too small, and the strong heating at the accretion shock (where the infalling envelope material meets the disc) acts to stabilise the disc.

This inference must be qualified as follows. (i) The core parameters used are very uncertain, in particular the core masses. (ii) Many of the cores appear already to have formed protostars (as evidenced by their association with IRAS sources), and even those that have not may already be in an advanced state of collapse. Therefore the observed values of $R_{_{\rm O}}$ and $\Omega_{_{\rm O}}$ may not be ``initial'' in the sense we have defined them. (iii) Our simulations do not take account of the low levels of turbulence observed in such cores. (iv) We do not rule out the possibility that at a later stage, once infall ceases, some of these protostellar discs cool and fragment, particularly if more efficient cooling due to dust operates.

\section{Conclusions}

We have performed simulations of the collapse of rigidly rotating prestellar cores. Each core is initially a mildly supercritical Bonnor-Ebert sphere ($\xi_{_{\rm B}}=6.9$), with its density increased by 10\%, and contained by external pressure. The gas is pure molecular hydrogen with  $T=28\,{\rm K}$ (hence isothermal sound speed $0.34\times 10^5\,{\rm cm}\,{\rm s}^{-1}$), the initial central density is $\rho_{_{\rm C}}=10^{-18}\,{\rm g}\,{\rm cm}^{-3}$, and hence the core mass is $M_{_{\rm O}}=6.1\,{\rm M}_\odot$, its initial radius is $R_{_{\rm O}}=17,000\,{\rm AU}$, and its external pressure is $P_{_{\rm EXT}}=k_{_{\rm B}}\,5.5\times 10^5\,{\rm cm}^{-3}\,{\rm K}$. The initial angular speed is varied from $\Omega_{_{\rm O}}=0.6\times 10^{-13}\,{\rm s}^{-1}$ to $\Omega_{_{\rm O}}=1.8\times 10^{-13}\,{\rm s}^{-1}$ (corresponding to $0.013<\beta<0.12$).

A slowly rotating core collapses to form a relatively massive primary protostar, and then a protostellar disc grows around this protostar.The disc is massive, but quite hot and compact, and so it does not fragment; it simply accretes onto the primary protostar. There is some observational evidence for the existence of massive discs \citep[e.g.][]{Eisner2005, Rodriguez2005, Greaves2008, Mann2009}. However, the phase we have simulated is short lived and still deeply embedded in the parental core, so our discs would be rather hard to observe.
We conjecture that such discs will show signatures of crystalline silicates out to disc radii of several AU.

A rapidly rotating core collapses initially to form a protostellar disc, and then a protostar condenses out at the centre. The protostar acquires almost all its mass via the disc, and therefore it grows quite slowly. The disc is extended, cool and thin, and tends to fragment.

On the basis of an analytic model, we obtain a condition on the mass, initial radius and initial angular speed of a core (Eqn. \ref{EQN:MULT}) such that its protostellar disc will fragment, whilst material is still infalling from the core envelope. Applying this condition to observed samples of cores we predict that such fragmentation is unlikely.

In a companion paper (Walch et al., in preparation), we explore how these results are fundamentally changed if a prestellar core is turbulent, but has the same net angular momentum as the cores modeled here.

\section*{Acknowledgments}
We thank D. Neufeld for providing the cooling tables. S. Walch performed this work with support from the International Max-Planck Research School, the DFG Cluster of Excellence {\it Origin and Structure of the Universe} (www.universe-cluster.de) and the Marie Curie RTN {\sc Constellation} (MRTN-CT-2006-035890). We thank the referee for his/her detailed and thoughtful comments, which were very helpful.
\bibliographystyle{mn2e}
\bibliography{references}

\appendix
\section[]{Analytic model}

We develop here a simple model to explain why rapidly rotating cores tend to fragment (and thereby presumably produce multiple systems), whereas slowly rotating cores do not. For simplicity we assume that the initial core is cylindrical, with height equal to its diameter and hence initial density $\rho_{_{\rm O}}=M_{_{\rm O}}/2\pi R_{_{\rm O}}^3$; we make this assumption purely because it allows us to formulate analytically the structure of the resulting disc. The core is initially in solid-body rotation with angular speed $\Omega_{_{\rm O}}$, and we assume that during the Isothermal Collapse Phase, {\it and} during the early Protostellar Disc Phase, angular momentum is conserved minutely. Consequently the material initially in an annulus at radius $r_{_{\rm O}}$ has mass $M_{r_{\rm o}}=M_{_{\rm O}}r_{_{\rm O}}^2/R_{_{\rm O}}^2$ ``interior to it'' in the disc, and ends up at radius $r$ with angular speed $\Omega_{_{\rm D}}(r)=\Omega_{_{\rm O}}r_{_{\rm O}}^2/r^2$. Centrifugal balance then requires that $GM_{r_{\rm o}}/r^2\simeq\Omega_{_{\rm D}}^2(r)r$, and hence 
\begin{eqnarray}
r\!\!&\!\!\simeq\!\!&\!\!\frac{R_{_{\rm O}}^2\Omega_{_{\rm O}}^2r_{_{\rm O}}^2}{GM_{_{\rm O}}}\,.
\end{eqnarray}
The edge of the disc is therefore at
\begin{eqnarray}
R_{_{\rm D}}\!\!&\!\!\simeq\!\!&\!\!\frac{R_{_{\rm O}}^4\Omega_{_{\rm O}}^2}{GM_{_{\rm O}}}\,,
\end{eqnarray}
and
\begin{eqnarray}
\Omega_{_{\rm D}}(r)\!\!&\!\!\simeq\!\!&\!\!\frac{G^2M_{_{\rm O}}^2}{R_{_{\rm O}}^6\Omega_{_{\rm O}}^3\hat{r}}\,,\hspace{0.7cm}\hat{r}\equiv\frac{r}{R_{_{\rm D}}}.
\end{eqnarray}
It follows that the final surface density of the disc is
\begin{eqnarray}
\Sigma_{_{\rm D}}(r)\!\!&\!\!=\!\!&\!\!\frac{M_{_{\rm O}}}{\pi R_{_{\rm O}}^2}\,\frac{2\pi r_{_{\rm O}}dr_{_{\rm O}}}{2\pi rdr}\;\simeq\;\frac{G^2M_{_{\rm O}}^3}{2\pi R_{_{\rm O}}^8\Omega_{_{\rm O}}^4\hat{r}}.
\end{eqnarray}
Infalling material impinges on the disc at radius $r$ with speed
\begin{eqnarray}
v_{_{\rm D}}(r)\!\!&\!\!\simeq\!\!&\!\!\left(\frac{GM_{r_{\rm o}}}{r}\right)^{\frac{1}{2}}\;\simeq\;\frac{GM_{_{\rm O}}}{R_{_{\rm O}}^2\Omega_{_{\rm O}}}\,.
\end{eqnarray}
We shall assume that it starts arriving at time
\begin{eqnarray}
t_{_{\rm D}}(r)\!\!&\!\!\simeq\!\!&\!\!\left(\!\frac{r_{_{\rm O}}^3}{GM_{r_{\rm o}}}\!\right)^{\frac{1}{2}}\,\simeq\,t_{_{\rm FF}}\hat{r}^{\frac{1}{4}},\hspace{0.7cm}t_{_{\rm FF}}\,=\,\left(\!\frac{R_{_{\rm O}}^3}{GM_{_{\rm O}}}\!\right)^{\frac{1}{2}},
\end{eqnarray}
and continues arriving at a constant rate until time $2t_{_{\rm D}}(r)$. Hence, in the  time interval $t_{_{\rm D}}(r)\leq t\leq 2t_{_{\rm D}}(r)$, the flux of material onto one side of the disc at radius $r$ is
\begin{eqnarray}
\Phi_{_{\rm D}}(r)\!\!&\!\!\simeq\!\!&\!\!\frac{\Sigma_{_{\rm D}}(r)}{2t_{_{\rm D}}(r)}\,\simeq\,\left(\!\frac{GM_{_{\rm O}}}{R_{_{\rm O}}^3}\!\right)^{\!\frac{5}{2}}\frac{M_{_{\rm O}}}{4\pi R_{_{\rm O}}^2\Omega_{_{\rm O}}^4\hat{r}^{5/4}};
\end{eqnarray}
the instantaneous surface-density of the disc is
\begin{eqnarray}\nonumber
\Sigma(r,t)\!\!&\!\!=\!\!&\!\!\Sigma_{_{\rm D}}(r)\left\{\frac{t}{t_{_{\rm D}}(r)}-1\right\}\\
&=&\!\!\frac{G^2M_{_{\rm O}}^3}{2\pi R_{_{\rm O}}^8\Omega_{_{\rm O}}^4\hat{r}}\left\{\frac{\hat{t}}{\hat{r}^{\frac{1}{4}}}-1\right\},\hspace{0.7cm}\hat{t}\equiv\frac{t}{t_{_{\rm FF}}}\,;
\end{eqnarray}
the ram pressure of the infalling material is
\begin{eqnarray}\nonumber\label{EQN:PRAM}
P_{_{\rm RAM}}(r)\!\!&\!\!\simeq\!\!&\!\!\Phi_{_{\rm D}}(r)v_{_{\rm D}}(r)\,\simeq\,\left(\!\frac{GM_{_{\rm O}}}{R_{_{\rm O}}^3}\!\right)^{\!\frac{7}{2}}\frac{M_{_{\rm O}}}{4\pi R_{_{\rm O}}\Omega_{_{\rm O}}^5\hat{r}^{\frac{5}{4}}};\\\label{EQN:PRAM}
\end{eqnarray}
and the flux of energy carried by the infalling material, and dissipated in the accretion shock, is
\begin{eqnarray}
F_{_{\rm D}}(r)\!\!&\!\!\simeq\!\!&\!\!\frac{\Phi_{_{\rm D}}(r)v_{_{\rm D}}^2(r)}{2}\,\simeq\,\left(\!\frac{GM_{_{\rm O}}}{R_{_{\rm O}}^3}\!\right)^{\!\frac{9}{2}}\frac{M_{_{\rm O}}}{8\pi\Omega_{_{\rm O}}^6\hat{r}^{\frac{5}{4}}}.
\end{eqnarray}
The disc cools mainly by molecular-line radiation. Under the conditions obtaining in the discs simulated here, the lines tend to be thermalised and optically thick. Hence the cooling rate per unit area of the disc, due to one species, M, is
\begin{eqnarray}\nonumber
F_{_{\rm M}}(r,t)\!\!&\!\!\sim\!\!&\!\!\frac{2^4\pi^{12}I_{_{\rm M}}k_{_{\rm B}}^9T^9(r,t)}{3^2A_{_{\rm M}}Z_{_{\rm M}}h^8c^6\Sigma(r,t)}\\
&\sim&\frac{2^5\pi^{13}I_{_{\rm M}}\left(k_{_{\rm B}}T\right)^9R_{_{\rm O}}^8\Omega_{_{\rm O}}^4\hat{r}}{3^2A_{_{\rm M}}Z_{_{\rm M}}h^8c^6G^2M_{_{\rm O}}^3}\left\{\frac{\hat{t}}{\hat{r}^{\frac{1}{4}}}-1\right\}^{-1},
\end{eqnarray}
where $I_{_{\rm M}}$ is the moment of inertia of M, $A_{_{\rm M}}$ is the Einstein-A coefficient of the $J=1\rightarrow 0$ transition of M, $Z_{_{\rm M}}$ is the fractional abundance of M (by mass). Since we are neglecting redistribution of angular momentum, we also neglect viscous heating of the disc; this is a reasonable assumption, given that the disc grows faster than the primary protostar (i.e. the viscous evolution of the disc proceeds on a longer time-scale than infall onto the disc). Hence the principal heating mechanism is dissipation of kinetic energy in the accretion shock, and therefore we estimate the temperature by setting $F_{_{\rm D}}(r,t)\sim F_{_{\rm M}}(r,t)$. It follows that, during the period when the disc is growing,
\begin{eqnarray}\nonumber
\left(k_{_{\rm B}}T(r,t)\right)^9\!\!&\!\!\sim\!\!&\!\!\left(\!\frac{GM_{_{\rm O}}}{R_{_{\rm O}}^3}\!\right)^{\!\frac{13}{2}}\frac{3^2A_{_{\rm M}}Z_{_{\rm M}}h^8c^6M_{_{\rm O}}^2}{2^8\pi^{14}I_{_{\rm M}}R_{_{\rm O}}^2\Omega_{_{\rm O}}^{10}\hat{r}^{\frac{9}{4}}}\left\{\frac{\hat{t}}{\hat{r}^{\frac{1}{4}}}-1\right\};\\
\end{eqnarray}
The Toomre and Gammie parameters for the actively accreting part of the disc are
\begin{eqnarray}
Q_{_{\rm T}}(r,t)\!\!&\!\!\equiv\!\!\!&\!\!\!\frac{a(r,t)\Omega_{_{\rm D}}(r)}{\pi G\Sigma(r,t)}\,,\\
Q_{_{\rm G}}(r,t)\!\!&\!\!\equiv\!\!\!&\!\!\!\frac{t_{_{\rm COOL}}(r,t)}{t_{_{\rm DYN}}(r,t)}\,=\,\frac{\pi G\Sigma^2(r,t)a(r,t)}{24F_{_{\rm D}}(r)}
\end{eqnarray}
To proceed, we consider cooling by CO, for which $I_{_{\rm CO}}\simeq 1.4\times 10^{-39}\,{\rm g}\,{\rm cm}^2$, $A_{_{\rm CO}}\simeq 10^{-7}\,{\rm s}^{-1}$, and $Z_{_{\rm CO}}\simeq 0.002$; and we introduce dimensionless core parameters,
\[
\hat{M}_{_{\rm O}}\equiv\frac{M_{_{\rm O}}}{6.1\,{\rm M}_\odot}\,,\;\;\;\;\;\hat{R}_{_{\rm O}}\equiv\frac{R_{_{\rm O}}}{17,000\,{\rm AU}}\,,\;\;\;\;\;\hat{\Omega}_{_{\rm O}}\equiv\frac{\Omega_{_{\rm O}}}{10^{-13}\,{\rm s}^{-1}}\,.
\]
Then,
\begin{eqnarray}
T(r,t)\!\!&\!\!\simeq\!\!&\!\!60{\rm K}\;\hat{M}_{_{\rm O}}^{\frac{17}{18}}\,\hat{R}_{_{\rm O}}^{-\frac{43}{18}}\,\hat{\Omega}_{_{\rm O}}^{-\frac{10}{9}}\,\hat{r}^{-\frac{1}{4}}\,\left\{\frac{\hat{t}}{\hat{r}^{\frac{1}{4}}}-1\right\}^{\!\frac{1}{9}};\\
Q_{_{\rm T}}(r,t)\!\!&\!\!\simeq\!\!&\!\!0.76\;\hat{M}_{_{\rm O}}^{-\frac{19}{36}}\,\hat{R}_{_{\rm O}}^{\frac{29}{36}}\,\hat{\Omega}_{_{\rm O}}^{\frac{4}{9}}\,\hat{r}^{-\frac{1}{8}}\,\left\{\frac{\hat{t}}{\hat{r}^{\frac{1}{4}}}-1\right\}^{\!-\frac{17}{18}};\\\label{EQN:TOOMREQ2}
Q_{_{\rm G}}(r,t)\!\!&\!\!\simeq\!\!&\!\!0.19\;\hat{M}_{_{\rm O}}^{\frac{35}{36}}\,\hat{R}_{_{\rm O}}^{-\frac{133}{36}}\,\hat{\Omega}_{_{\rm O}}^{-\frac{23}{9}}\,\hat{r}^{-\frac{7}{8}}\,\left\{\frac{\hat{t}}{\hat{r}^{\frac{1}{4}}}-1\right\}^{\!\frac{37}{36}}.
\end{eqnarray}
The mass of the system (primary protostar plus protostellar disc) is given by
\begin{eqnarray}
\frac{M_{_{\rm SYS}}(t)}{M_{_{\rm O}}}\!\!&\!\!=\!\!&\!\!\left\{\!\!\begin{array}{ll}
\frac{7}{24}\hat{t}^4,&0\;\leq\! \hat{t}\!\leq 1;\\
\frac{4}{3}\hat{t} -1-\frac{1}{24}\hat{t}^4,\,\hspace{0.5cm}& 1\;<\!\hat{t}\!<2;\\
1,&2 \;\leq \!\hat{t}.
\end{array}\right.
\end{eqnarray}
It follows that at $t=t_{_{\rm FF}}$ (i.e. $\hat{t}=1$), $M_{_{\rm SYS}}\simeq 0.29M_{_{\rm O}}$, and since this is very close both to the juncture when the disc is most unstable, and to the juncture at which we compare simulations with different $\Omega_{_{\rm O}}$, we will evaluate the stability of the discs at this time too. At $t=t_{_{\rm FF}}$, the disc is complete out to $R_{_{\rm D}}/16$, and is just starting to form at $R_{_{\rm D}}$, so between these radii the disc is actively accreting. From Eqn. (\ref{EQN:TOOMREQ2}) we see that the accreting disc is most unstable (Toomre $Q_{_{\rm T}}\sim 1.07$) at $R_{_{\rm D}}/16$, and this is basically where the simulated discs with high $\Omega_{_{\rm O}}$ tend to fragment. The condition for the disc to be able to cool fast enough to fragment at $R_{_{\rm D}}/16$ (i.e. Gammie $Q_{_{\rm G}}<1$) then reduces to
\begin{eqnarray}\label{EQN:GAMMIEQ3}
\hat{\Omega}_{_{\rm O}}\!\!&\!\!>\!\!&\!\!1.35\,\hat{M}_{_{\rm O}}^{\frac{35}{92}}\,\hat{R}_{_{\rm O}}^{-\frac{133}{92}}\,.
\end{eqnarray}
Eqn. (\ref{EQN:GAMMIEQ3}) agrees with the numerical results, which give fragmentation for $\hat{\Omega}_{_{\rm O}}=1.5$ but not for $\hat{\Omega}_{_{\rm O}}=1.2$. It also indicates how the criterion for fragmentation scales with core mass and radius, and we exploit this to analyse observed cores in Section \ref{SEC:OBS}. The analytic model also accords well with the simulations in predicting that fragmentation should occur near $R_{_{\rm D}}/16$, which for those discs that fragment is between $\sim 200\,{\rm AU}$ and $\sim 700\,{\rm AU}$ \citep[cf.][]{Rafikov2005, Matzner2005,  Whitworth2006}.

\label{lastpage}
\end{document}